\documentclass[12pt,preprint]{aastex} 

\newcommand{\be}{\begin{equation}}
\newcommand{\ee}{\end{equation}}
\newcommand{\rot}{ {r_\omega} }
\newcommand{\mpl}{ \Gamma  } 
\newcommand{\sigwid}{{ {\hat \sigma} }}
\newcommand{\const}{{ c_{\rm e} }} 
\newcommand{\step}{{ \Theta }} 
\newcommand{\rexp}{{ R_{\rm x} }} 
\newcommand{\length}{{ \ell_\sigma }} 
\newcommand{\qam}{{ \eta }} 
\def\lta{\,\raise 0.3 ex\hbox{$ < $}\kern -0.75 em
 \lower 0.7 ex\hbox{$\sim$}\,}
\def\gta{\,\raise 0.3 ex\hbox{$ > $}\kern -0.75 em
 \lower 0.7 ex\hbox{$\sim$}\,} 

\begin{document} 

\title{BARYONIC COLLAPSE WITHIN DARK MATTER HALOS \\ 
AND THE FORMATION OF GASEOUS GALACTIC DISKS} 

\author{Fred C. Adams$^{1,2}$ and Anthony M. Bloch$^{1,3}$} 

\affil{$^1$Michigan Center for Theoretical Physics \\
Physics Department, University of Michigan, Ann Arbor, MI 48109} 

\affil{$^2$Astronomy Department, University of Michigan, Ann Arbor, MI 48109} 

\affil{$^3$Department of Mathematics, University of Michigan, Ann Arbor, MI 48109} 

\begin{abstract} 

This paper constructs an analytic framework for calculating the
assembly of galactic disks from the collapse of gas within dark matter
halos, with the goal of determining the resulting surface density
profiles.  In this formulation of the problem, gas parcels (baryons)
fall through the potentials of dark matter halos on nearly ballistic,
zero energy orbits and collect in a rotating disk. The dark matter
halos have a nearly universal form, as determined previously through
numerical simulations. The simplest scenario starts with a gaseous
sphere in slow uniform rotation, follows its subsequent collapse, and
determines the surface density of the disk. This calculation is
carried out for pre-collapse mass distributions of the form $M(r) \sim
r^p$ with $1 \le p \le 3$ and for polytropic spheres in hydrostatic
equilibrium with the halo potential.  The resulting disk surface
density profiles have nearly power-law forms $\sigma \sim \varpi^{-q}$
(where $\varpi$ is the radial coordinate in the disk), with
well-defined edges determined by the centrifugal barrier $R_C$ -- the
radius to which gas with the highest specific angular momentum falls
during the collapse. This idealized scenario is generalized to include
non-spherical starting states, alternate rotation profiles, and
multiple accretion events (e.g., due to gas being added to the halo
via merger events). This latter complication is explored in some
detail and considers a log-normal distribution for the angular momenta
of the pre-collapse states of the individual components. In the limit
where this distribution is wide, the composite surface density
approaches a universal form $\sigma_T \sim \varpi^{-2}$, independent
of the shape of the constituent profiles. When the angular momentum
distribution has an intermediate width, however, the composite surface
density attains a nearly exponential form, roughly consistent with
profiles of observed galaxies.

\end{abstract}


\section{INTRODUCTION} 

Although the formation of dark matter halos is rapidly becoming
understood, the formation of gaseous galactic disks remains more
elusive.  Disk formation is complicated and involves a number of
physical processes, including the large scale collapse of the dark
matter, merging of separate halos, gas cooling, and the infall of gas
through the gravitational potential of the dark matter. Numerical
simulations now include all of these processes and are beginning to
produce realistic gaseous disks (e.g., Governato et al. 2004;
Robertson et al. 2004; Keres et al. 2005).  This paper adopts a
complementary approach and seeks to study the infall process ---
separately and analytically --- where the starting conditions are
informed by numerical simulations that have been previously carried
out. This paper thus develops an analytic framework from which to
study the infall process in galactic disk formation. This formalism
applies to a wide range of infall scenarios and can be readily
generalized to include additional physical processes. The main focus
of this paper is to investigate the range of surface density profiles
produced by gaseous collapse within the potential of a dark matter
halo.

Although the aforementioned simulations demonstrate that the current
paradigm of structure formation can produce galactic disks with
reasonable properties, an analytic description of disk formation is
useful in several respects. Analytic solutions provide us with a
direct assessment of how the final disk surface density properties
depend on starting conditions and other parameters of the problem.
Along the way, we obtain an analytic description of several ancillary
quantities, e.g., the velocity and density fields of the gaseous
material as it falls toward the disk, as well as the hydrostatic
equilibrium profiles of the pre-collapse states.  Once found, the
resulting analytic solutions, both for the ancillary quantities and
the resulting disk surface density profiles, can be used in a variety
of other contexts.

In the standard picture of galactic disk formation (e.g., Rees \&
Ostriker 1977; Fall \& Efstathiou 1980; White \& Frenk 1991), the gas
within a forming dark matter halo approaches a density profile that is
roughly parallel to that of the dark matter.  This assumption is
justified by complementary simulations of gas dynamics (from Evrard et
al. 1994 to Kaufmann et al. 2006). In order for the gas to fall inward
toward the nascent galactic disk, the gas must cool.  The gas cooling
time depends primarily on density, which is a decreasing function of
galactic radius, so the cooling time increases outward (note that
metallicity, temperature, etc., also influence the cooling time).
There are thus two regimes of interest. In one limit, the entire
virialized region has a cooling time shorter than the age of the
universe (at the formation time of the halo).  In this case, the gas
cools rapidly and never approaches hydrostatic equilibrium. The gas
supply is then limited by the rate at which the forming galactic
structure accretes gas from larger radii. In the other limit, the
cooling time is longer than the age of the universe for much of the
virialized region, and infall is limited by the rate of gas cooling
(White \& Rees 1978). The flow of gas into the central regions is then
analogous to cooling flows (e.g., Fabian et al. 1984). In both
regimes, the gas falls inward through the gravitational potential
provided (primarily) by the dark matter halo.

In this paper, we study the infall of baryonic gas moving in the
potential of a dark matter halo.  Since dark matter halos approach a
known nearly universal profile (e.g., Navarro et al. 1997; hereafter
NFW), we consider the potential to have a given form and follow the
orbits of gas parcels as they fall through the dark matter and build
up a galactic disk structure.  Specifically, we consider the halo to
have a fixed profile, independent of the gas, and to take the form of
a Hernquist potential (Hernquist 1990; eq. [\ref{eq:hq}]). Notice that
this form is essentially the same as the more commonly used NFW
profile, except that it has a steeper density profile at large radii
and hence converges to a finite mass (see also Busha et al. 2005). As
outlined above, the gas must cool before it falls through the halo
potential toward the central regions. For the first part of this work,
we describe this process as a single, overall collapse, even though
the gas cools at different rates.  This complication will not affect
the resulting surface density profiles of the disk as long as the
central portions fall in first (as expected, since they tend to be
denser and cool faster) so that infalling shells do not cross.  Notice
also that we are implicitly assuming that the gas mass is small
compared that of the dark matter.  After a sufficiently large quantity
of baryonic mass has accumulated near the galactic center, however,
the back reaction of the gas on the dark matter potential should be
included.

The goal of this calculation is thus relatively modest: We want to
develop an analytic transformation between starting gas distributions
and the ``final'' surface density profiles of the disk. In this
context, the starting gas distributions reflect the initial profiles
of specific angular momentum, which depend on gas cooling and other
effects that determine the starting state. The final surface density
profiles (as considered here) are determined by where gas falls onto
the galactic disk with no additional evolution (e.g., due to
gravitational torques, which act on longer time scales). We discuss
the effects of additional evolution, e.g., angular momentum
redistribution, in the final section.

This work expands upon previous work concerning galactic disk
formation (see, e.g., Gunn 1982, Kauffmann 1996, Dalcanton et al.
1997, Firmani \& Avila-Reese 2000, Bullock et al. 2001, van den Bosch
2001, Thacker \& Couchman 2001). These previous studies explore disk
formation within dark matter halos with increasing levels of
complexity, primarily through numerical simulations. The generic
finding is that the resulting surface density profiles are more
concentrated in the center than an exponential law, and typically
display sharper outer boundaries.  This result depends on the starting
(pre-collapse) distribution of angular momentum of the gas.  Dalcanton
et al. (1997) start with a uniformly rotating sphere, Firmani \&
Avila-Reese (2000) assume that each infalling mass shell has a
constant spin parameter $\lambda$, and Bullock et al. (2001) determine
the initial angular momentum distribution directly from N-body
simulations. In these studies, the disk surface density profiles are
the result of angular momentum conservation, although the initial
distributions of angular momentum vary. One contribution of this
present study is to provide a systematic exploration of possible
starting states and their effect on the corresponding surface density
profiles. In particular, we generalize the starting conditions to
include a wide variety of initial density profiles, rotation profiles,
and initial geometries. We also explicitly determine the velocity
fields and density fields of the infalling material.  More
significantly, this analytic framework is sufficiently robust to
account for multiple accretion events (see also Firmani \& Avila-Reese
2000). For comparison, previous numerical work (Robertson et al. 2004;
Keres et al. 2005) includes more physical processes (e.g., explicit
calculation of gas cooling and angular momentum transport during
collapse), but does not isolate the dynamical mechanism(s) by which
galactic disks are assembled.

This paper is organized as follows. We first extract the relevant
orbit solutions that describe gas parcels falling through the
gravitational potential of a dark matter halo (\S 2); this calculation
determines the velocity and density fields of the infalling material,
as well as the locations where given parcels fall onto the disk plane.
In \S 3 we apply these results to the collapse of rotating gaseous
spheres and find the resulting surface density profiles as a function
of the starting density and angular momentum distributions. These
results are generalized (in \S 4) to include more complex starting
geometries, including holes, removal of polar caps, and filamentary
structures.  Motivated by the fact that cosmic structure is built up
through a hierarchical process, \S 5 considers the effects of multiple
accretion events and finds the corresponding composite surface density
profiles. The paper concludes in \S 6 with a summary of our results
and a discussion of how exponential disks can be produced.

\section{DYNAMIC INFALL IN A DARK MATTER HALO POTENTIAL} 

For a given gravitational potential, we must find orbital solutions
for gaseous material falling towards the galactic center. Here we
assume that the halo has the form of a Hernquist profile so that the
potential, density distribution, and mass profile are given by
\be 
\Psi = {\Psi_0 \over 1 + \xi} \, , \qquad \rho = 
{\rho_0 \over \xi (1 +  \xi)^3}, \qquad {\rm and} \qquad 
M = M_\infty {\xi^2 \over (1 + \xi)^2} \, , 
\label{eq:hq} 
\ee 
where $\xi = r/r_s$ and where $r_s$ is the scale length of the
structure.  We expect $r_s \sim 50 - 80$ kpc for a typical galactic
halo. The other constants are related to each other through the
expressions $\Psi_0 = 2 \pi G \rho_0 r_s^2$ and $M_\infty = 2 \pi
r_s^3 \rho_0$. Notice that $\Psi_0$ is defined to be positive (so that
the signs in the equations of motion must be chosen accordingly).
Notice also that the halo parameters $(\Psi_0, r_s, M_\infty)$ will 
evolve with time and will, in general, be different for different
accretion events. As a starting approximation, however, these
quantities are taken to be constant during a given accretion event.

We must specify the initial conditions for the infalling baryonic gas.
For any orbit, the initial condition can be specified by the energy
and angular momentum.  As first approximation, we consider the gas
parcels to enter the central region of the galaxy on zero energy
orbits. This assumption could be relaxed to include small but nonzero
energies, which can be written in dimensionless form
\be 
\epsilon = \, |E| / \Psi_0 \, \ll 1 \, . 
\ee 
In addition, the gas parcels are assumed to have a given specific
angular momentum $j = j_\infty$ at their starting radius. The angular
velocity vector of the starting (spherical) state points along the
$\hat z$ axis, but the angular momentum vector of an individual gas
parcel depends on its polar angle (see below).  In order for the gas
to fall relatively close to the galactic center -- onto the forming
galactic disk -- this angular momentum must be sufficiently small. 
Following previous work (Adams \& Bloch 2005; hereafter AB05), the
orbits are described by the angular momentum parameter
\be 
\qam \equiv j^2 / 2 \Psi_0 r_s^2 \, . 
\ee
The condition that the orbits fall in the central region of the 
halo thus takes the form $\qam \ll 1$. 

\begin{figure}
\figurenum{1}
{\centerline{\epsscale{0.90} \plotone{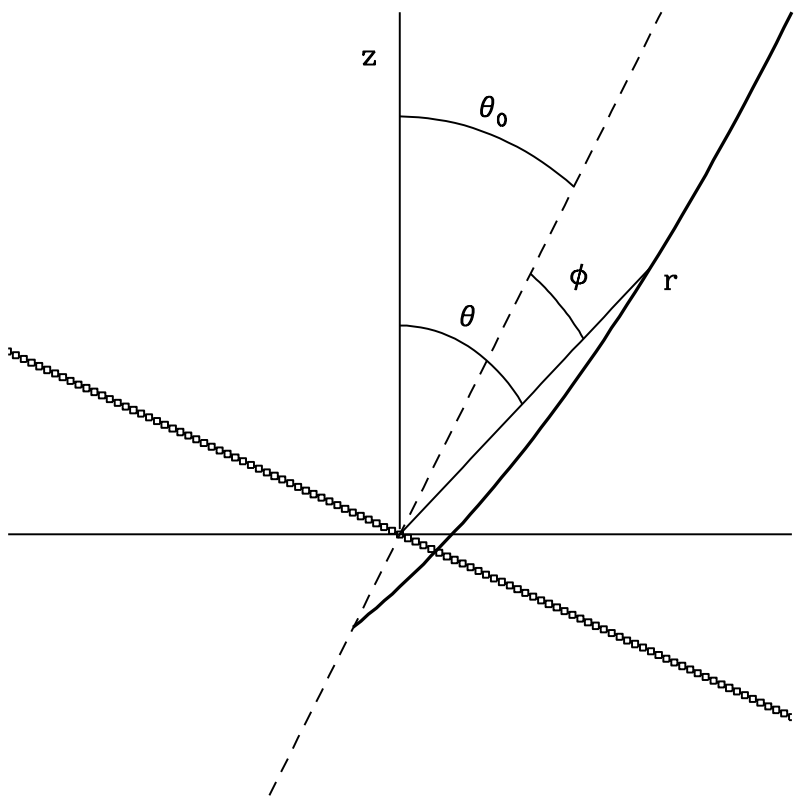} }}
\figcaption{Schematic diagram showing the geometry of the orbits. The
heavy solid curve depicts the orbit trajectory, which begins at large
radius $r_0 \gg r$ and continues inward until it reaches the disk
plane.  The orbit proceeds within a fixed plane that lies at an angle
$\theta_0$, measured with respect to the ${\hat z}$ axis (${\hat r}_0
\cdot {\hat z} = \cos \theta_0$). At a given point along the orbit,
marked by position vector ${\bf r} = r {\hat r}$, the angle within the
orbital plane is $\phi$ (${\hat r} \cdot {\hat r}_0 = \cos \phi$) and
the position angle in the original coordinate system is $\theta$
(${\hat r} \cdot {\hat z} = \cos \theta$). The thick line marked by 
square symbols shows the intersection of the orbit plane with the 
disk plane. } 
\label{fig:geometry} 
\end{figure}

In the limit of interest, $\qam \ll 1$ and $\epsilon \to 0$, the orbit 
solutions take the approximate form  
\be 
\xi \sin\phi = \sqrt{\qam} = {\it constant} \, , 
\label{eq:orbit0} 
\ee
where $\phi$ is the turning angle in the plane of the orbit (AB05).
Since the potential is spherically symmetric, angular momentum is
conserved and the motion is confined to a plane described by the
coordinates $(r, \phi)$; the radius $r$ is the same in both the plane
and the original spherical coordinates. The angular coordinate $\phi$
in the orbital plane is not the same as the polar angle $\theta$ in
the original spherical coordinates, but the two are simply related.
Suppose that the gas parcel has polar angle $\theta_0$ at the start of
its orbit, i.e., we let $\theta_0$ denote the angle of the
asymptotically radial streamline. The coordinate position of the gas
parcel is subsequently described by either the angle $\phi$ in the
orbital plane, or by the polar angle $\theta$ in the original
spherical coordinates (see Figure \ref{fig:geometry}). The two angles
obey a geometrical relation (e.g,. Ulrich 1976) that depends on the
initial polar angle, i.e.,
\be
\cos \phi = { \cos \theta \over \cos \theta_0} \equiv 
{\mu \over \mu_o} \, , 
\ee 
where the second equality defines $\mu \equiv \cos\theta$ and $\mu_0
\equiv \cos\theta_0$.  Combining the above relations, we can write the
orbit equation (\ref{eq:orbit0}) in terms of the variables $\mu$ and 
$\mu_0$ to obtain
\be
1 - {\mu^2 \over \mu_0^2} = {\qam \over \xi^2} \, . 
\label{eq:orbit} 
\ee 

For zero energy orbits, conservation of energy enforces a constraint
on the velocity fields of the form
\be 
v_r^2 + v_\theta^2 + v_\varphi^2 = {2 \Psi_0 \over 1 + \xi} \, . 
\ee 
Given the orbital solution, we can find the velocity fields 
\be
v_r^2 = 2 \Psi_0 
\Bigl\{ {1 \over 1 + \xi} - {\qam \over \xi^2} \Bigr\} \, , \qquad  
v_\theta^2 = 2 \Psi_0 \, \Bigl( {\qam \over \xi^2} \Bigr) \,  
{ \mu_0^2 - \mu^2 \over 1 - \mu^2 } \, , \qquad {\rm and} \qquad 
v_\varphi^2 = 2 \Psi_0 \, \Bigl( {\qam \over \xi^2} \Bigr) \,   
{  1 - \mu_0^2 \over 1 - \mu^2} \, . 
\label{eq:vfield} 
\ee 
Since $\mu$ and $\mu_0$ are related through the orbit equation
(\ref{eq:orbit}), the velocity field is completely determined for any
position $(r, \theta)$.  The density distribution of the infalling
material can be obtained by applying conservation of mass along a
stream-tube (Ulrich 1976; Chevalier 1983), i.e.,
\be 
\rho(r,\theta) \, v_r \, r^2
\sin\theta \, d\theta \, d\varphi = - { {\dot M} \over 4 \pi}
\sin\theta_0 \, d\theta_0 \, d\varphi_0 , 
\ee 
where $\dot M$ is the total rate of mass flow through a spherical
surface (note that $\dot M$ is defined in the intermediate asymptotic
regime inside the starting radii and outside the galactic disk). The
minus sign arises because the radial velocity $v_r$ is negative for
infalling trajectories.  Combining the above equations allows us to
write the density profile of the incoming material in the form
\be 
\rho(r,\theta) = { {\dot M} \over 4 \pi |v_r| r^2 }
{d \mu_0 \over d \mu} \, . 
\label{eq:density0} 
\ee
The properties of the collapsing structure determine the orbit
equation (\ref{eq:orbit}), which in turn determines the form of
$d\mu_0/d\mu$ (see also Cassen \& Moosman 1981; Terebey et al. 1984).

With the density specified, we can determine the rate at which the
forming galactic disk gains mass, $d\sigma/dt$ = $\rho v_\theta$,
where the right hand side of the equation is evaluated at the disk
plane $\theta$ = $\pi/2$ or $\mu$ = 0. Notice that we are assuming
that the newly arriving gaseous material dissipates the energy
contained within its $\hat r$ and $\hat \theta$ velocity components,
so that it joins the disk. Such dissipation can take place, e.g., by
shocks with the pre-existing disk material, or, in the case of
sufficiently symmetric flow, by meeting the corresponding gas parcel
arriving at the disk plane from the opposite direction. With these 
assumptions, the disk grows according to 
\be 
d \sigma = {d M \over 4 \pi r^2} \Bigg| {v_\theta \over v_r} \Bigg| 
{d \mu_0 \over d\mu} \, . 
\ee 

Notice that the above equation only accounts for the surface density
distribution produced by the arrival of the gas. After its arrival,
the gas will dissipate further energy, transfer angular momentum, and
the disk will evolve. In particular, the newly arriving material does
not know in advance that it will become part of a centrifugally
supported disk structure. As a result, it will not necessarily have
the proper azimuthal velocity appropriate for the local circular speed
of the disk (e.g., Cassen \& Moosman 1981). As it adjusts to disk
conditions, the gas will dissipate additional energy and move deeper
into the gravitational potential well of the galaxy. On longer time
scales, the disk will experience gravitational torques, which transfer
angular momentum and act to spread out the disk (e.g., Shu 1990;
Binney \& Tremaine 1987). These effects will be discussed in greater
detail below (see \S 6.2).

\section{BASIC COLLAPSE SOLUTIONS} 

In the standard picture of galactic disk formation, baryonic matter
initially traces the density distribution of the dark matter halo and
attains a similar distribution of angular momentum.  After cooling,
the gas falls inward on nearly ballistic (pressure-free) trajectories
and conserves its angular momentum. The objective of this work is to
understand this infall process in greater detail, starting form simple
models and then generalizing to more realistic models. In this section
we consider the surface density profiles produced through the collapse
of a single, well-defined gaseous configuration embedded within the
potential of the dark matter halo. We consider more general collapse
scenarios, including non-spherical effects and multiple accretion
events, in the following sections.

To start, we must thus specify the starting mass profile of the
(baryonic) gas.  Here we assume that the mass profile $M(r)$ is
spherically symmetric and that the inner regions fall before the outer
regions so that infalling shells do not cross. We also assume that the
initial structure is slowly rotating with a well-defined rotation
profile $\Omega (r)$.  Note that when a given mass $M$ has fallen to
the disk, the inverse relation $r(M)$ determines the initial location
-- and hence the angular momentum -- of the newly arriving material.
The mass profile $M(r)$ thus helps set the angular momentum profile of
the starting state for a given $\Omega (r)$, and thereby influences
the resulting surface density profile.

If the gas has an initially spherically symmetric density distribution
and is rotating with angular velocity $\Omega$, as described above,
the specific angular momentum for a given starting position
$(\xi_\infty, \theta_0)$ is given by
\be 
j_\infty = r_s^2 \xi_\infty^2 \Omega \sin \theta_0 \, ,  
\ee 
which implies that the angular momentum parameter takes the form 
\be 
\qam = \, {r_s^2 \Omega^2 \over 2 \Psi_0} \, \, 
\xi_\infty^4 \, (1 - \mu_0^2) \, . 
\ee
The orbit equation (\ref{eq:orbit}) thus becomes 
\be 
1 - {\mu^2 \over \mu_0^2} = {\qam_0 \over \xi^2} (1 - \mu_0^2) \, , 
\label{eq:monorbit} 
\ee
where we have defined 
\be
\qam_0 \equiv {r_s^2 \Omega^2 \over 2 \Psi_0}  \xi_\infty^4 
\equiv \omega^2 \xi_\infty^4 \, , 
\label{eq:qzero} 
\ee
where the second equality defines the dimensionless rotation  
parameter $\omega$. 

With the above considerations, the density field is thus completely
determined in analytic but implicit form 
\be 
\rho(\xi, \theta) = { {\dot M} \over 4 \pi r^2 \sqrt{2 \Psi_0} } 
{1 - \mu_0^2 \over \xi^2 \mu_0^2} \, .  
\label{eq:density1} 
\ee
Notice that the density must be expressed in terms of spherical
coordinates $r$ and $\theta = \cos^{-1} \mu$, but the right hand side
of the equation is written in terms of $\mu_0 = \cos \theta_0$. The
two variables are related through the orbit equation
(\ref{eq:monorbit}).

For the case of spherical collapse, we can now evaluate $d\mu_0/d\mu$,
the velocity fields, the density profile, and hence the rate at which
the disk gains mass from the infall.  We thus obtain
\be 
d \sigma = \rho(\xi, \theta=\pi/2) v_\theta dt = 
{ d M \over 2 \pi r_s^2} {1 \over \qam_0 } 
\big[ 1 - \xi^2 / \qam_0 \bigr]^{-1/2} \, , 
\label{eq:dsigma} 
\ee 
where we have multiplied the right hand side by a factor of 2 to
account for gas flowing onto the disk from both sides (top and
bottom).  In this model, $\qam_0 \propto r_\infty^4$ so the the
starting mass profiles [basically, how the enclosed mass $M(r)$
depends on $r$] determine how much mass initially has a given specific
angular momentum and hence the radial location on the disk where it
will fall. Notice that the derivation of equation (\ref{eq:dsigma})
uses an extreme approximation in which the turning point of the orbit
occurs at the point $\mu$ = 0, i.e., where the orbit intersects the
disk. As a result, both the radial velocity $v_r$ and the differential
$d \mu_0 / d\mu$ contain integrable singularities, which must be
handled properly. As expected, these singularities cancel out and
reveal a finite mass flow onto the disk.

For each given mass and angular momentum profile, the above formalism
allows us to calculate the corresponding surface density of the
resulting disk (see \S 3.1 -- \S 3.5).  We begin with some general
definitions. For any given mass profile, one can define a mass scale
$M_s$ to be the mass in the gaseous sphere contained within the scale
radius $r_s$ of the dark matter halo, i.e.,
\be
M_s \equiv M(r_s) \, . 
\ee
This mass scale is a convenient reference mass. Notice that $M_s$ is
the baryonic mass and is thus a small fraction ($\sim 1/6$, Spergel et
al. 2003) of the total mass contained within the radius $r_s$.  We can
also define a rotational $\rot$ within the collapsing gaseous sphere, i.e.,  
\be
\rot \equiv r_s \omega = r_s^2 \Omega / (2 \Psi_0)^{1/2} \, . 
\ee 
As shown below, the centrifugal barrier, which sets the location of
the outer disk edge, is given by a dimensionless factor times the
scale $\rot$.  As gaseous material falls inward, it has a given radius
$r$ in spherical coordinates, and then becomes part of a disk which is
generally described in terms of cylindrical coordinates. The radius $r
= r_s \xi$ in spherical coordinates thus becomes the radius within the
disk in cylindrical coordinates and is denoted as
\be
\varpi \equiv r_s \xi \, . 
\ee 
With these definitions, we can find the surface density profiles 
for a given starting state. 

\subsection{Specific spherical starting states with constant rotation rate} 

We can consider a range of specific models for the starting mass
profile, including a self-gravitating isothermal sphere
\be 
\rho(r) = {a^2 \over 2 \pi G r^2} \, , \qquad 
M(r) = {2 a^2 \over G} r \, , \qquad {\rm and} \qquad 
r(M) = { G M \over 2 a^2 } \, , 
\label{eq:isosphere} 
\ee 
where $a$ is the isothermal sound speed. We also consider a uniform density sphere  
\be 
\rho(r) = \rho_0 \, , \qquad M(r) = {4 \pi \over 3} \rho_0 r^3 \, , \qquad 
{\rm and} \qquad r(M) = \Bigl( { 3 M \over 4 \pi \rho_0 } \Bigr)^{1/3} \, , 
\label{eq:unisphere} 
\ee 
and an intermediate density profile motivated by the Hernquist (and NFW) 
profile, where the (pre-collapse) density profile takes the form  
\be 
\rho(r) = \rho_0 \bigl( r_0 / r \bigr) \, , ,\qquad 
M(r) = 2 \pi \rho_0 r_0 r^2 \, , \qquad {\rm and} \qquad 
r(M) = \Bigl( {M \over 2 \pi \rho_0 r_0} \Bigr)^{1/2} \, . 
\label{eq:logasphere} 
\ee
This profile would occur, e.g., if the baryonic gas traces the density
distribution of the dark matter halo in the inner regime. For this
profile, without loss of generality, we can take $r_0$ = $r_s$ and
define the constant $\rho_0$ accordingly.

Starting with the case of the isothermal sphere
(eq. [\ref{eq:isosphere}]), we find 
\be
d \sigma = {d M \over 2 \pi r_s^2} {1 \over \omega^2 (M/M_s)^4} 
\bigl[ 1 - \xi^2 \omega^{-2} (M_s/M)^4 \bigr]^{-1/2} \, . 
\ee
This expression can be written in terms of the dimensionless variable
$x \equiv (M/M_s) (\omega/\xi)^{1/2}$, and hence the integral
expression for the surface density becomes
\be 
\sigma (\varpi) = {M_s \over 2 \pi \varpi^{3/2} \rot^{1/2} } \, 
\int_1^{x_f} {dx \over x^4} \bigl[ 1 - x^{-4} \bigr]^{-1/2} \, . 
\ee
The range of integration starts at $x=1$, rather than $x=0$, because 
all of the material with $x < 1$ falls inside of the radial coordinate 
$\varpi$ in the disk.  For a given physical radius $\varpi$ within the
disk, the innermost portion of the sphere has too little angular
momentum (even at the equator where angular momentum is maximum) to
fall as far out in the disk as the radial location $\varpi$. The upper
limit of integration is set by the total mass $M_T$ of the accretion
event, i.e., $x_f = (M_T/M_s) (\rot/\varpi)^{1/2}$. In this scenario, 
the disk has a well-defined outer edge determined by the material with 
the highest specific angular momentum in the initial state. We denote 
this radius as the centrifugal barrier $R_C$, which takes the form 
$R_C = \rot (M_T / M_s)^2$, so that $x_f = (R_C/\varpi)^{1/2}$. 
The surface density can thus be written in the form 
\be 
\sigma (\varpi) = {M_T \over 2 \pi \varpi^{3/2} R_C^{1/2} } \, 
\int_1^{x_f} {dx \over x^4} \bigl[ 1 - x^{-4} \bigr]^{-1/2} \, . 
\ee
Note that when $\varpi \ll R_C$, far from the outer disk edge, the
surface density attains a nearly power-law form (here $\sigma \sim
\varpi^{-3/2}$).  Near the edge, where $\varpi \to R_C$, the upper 
limit of the integral $x_f \to 1$ and the surface density $\sigma \to
0$.

A similar treatment can be used for the other mass profiles. For the
case of a uniform density sphere (eq. [\ref{eq:unisphere}]), we define
$x \equiv (M/M_s) (\omega/\xi)^{3/2}$, and find the surface density 
\be 
\sigma (\varpi) = {M_T \over 2 \pi \varpi^{1/2} R_C^{3/2} } \, 
\int_1^{x_f} {dx \over x^{4/3}} \bigl[ 1 - x^{-{4/3} } \bigr]^{-1/2} \, , 
\ee
where $R_C$ = $\rot (M_T/M_s)^{2/3}$ and hence $x_f$ = $(R_C/\varpi)^{3/2}$. 
For the intermediate case (eq. [\ref{eq:logasphere}]), we define 
$x \equiv (M/M_s) (\omega/\xi)$, and find the surface density  
\be 
\sigma (\varpi) = {M_s \over 2 \pi \varpi \rot } \, 
\int_1^{x_f} {dx \over x^{2}} \bigl[ 1 - x^{-{2} } \bigr]^{-1/2} \, = 
{M_T \over 2 \pi R_C \varpi} \, \cos^{-1} (\varpi/R_C) \, , 
\label{eq:hqsigma} 
\ee
where $M_T$ is the total mass that has fallen to the disk and 
$R_C = (M_T/M_s) \rot$.

\begin{figure}
\figurenum{2}
{\centerline{\epsscale{0.90} \plotone{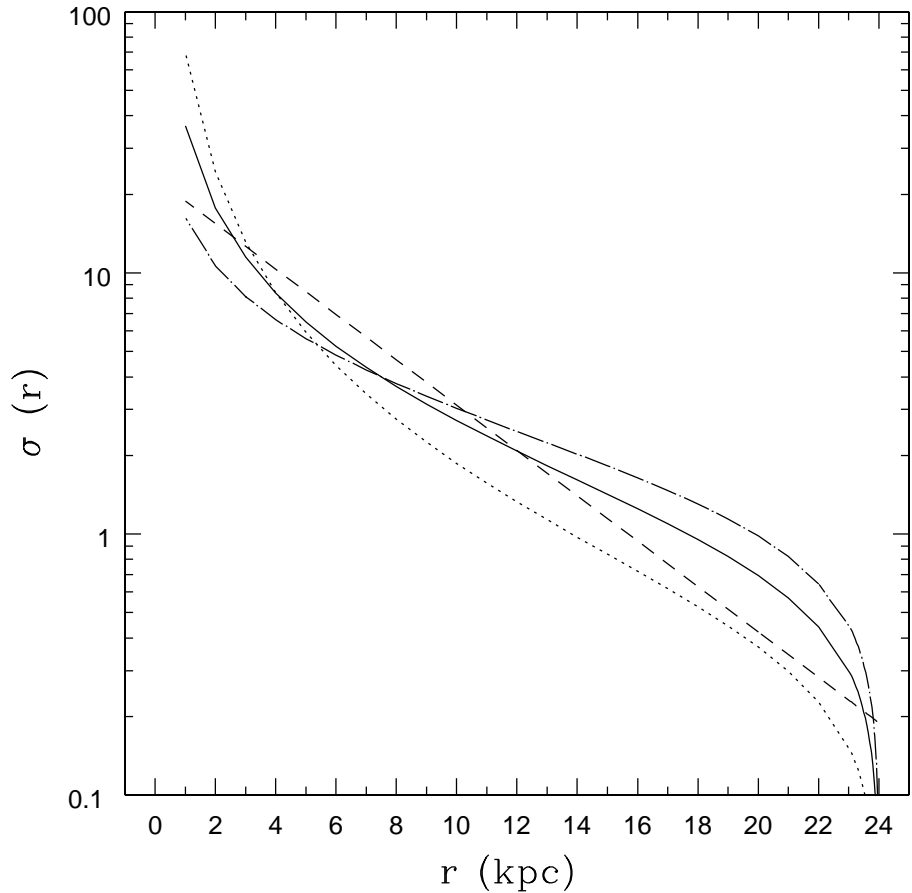} }}
\figcaption{Surface density profiles resulting from spherical 
collapse for three values of the index $p$ of the starting mass
profile $M(r) \propto r^p$: $p$ = 1 (dotted curve), $p$ = 2 (solid
curve), and $p$ = 3 (dot-dashed curve).  The rotation rate of the
starting states is taken to be a constant.  All three profiles have
the same total mass and the same value of the centrifugal barrier
(scaled here so that $R_C$ = 24 kpc), which determines the outer disk
edge.  The dashed curve shows an exponential profile for comparison,
where the exponential scale length is $\rexp$ = 5 kpc and the total
disk mass is the same as the other profiles. The scale on the vertical 
axis is arbitrary. }
\label{fig:profiles} 
\end{figure}

To illustrate the surface density profiles that result from a single
spherical collapse, Figure \ref{fig:profiles} shows the results for
three mass profiles $M(r) \sim r^p$, where $p$ = 1, 2, and 3. All
three profiles have the same total disk mass and the same centrifugal
radius, which sets the location of the outer disk edge. These profiles
are scaled so that $R_C$ = 24 kpc, but solutions can be found for any
disk mass and size $R_C$. To fix ideas, consider halo/disk parameters
roughly comparable to those of the Galaxy. If we use a Hernquist
profile to describe the Galactic potential, the scale length $r_s
\approx 65$ kpc (see AB05 for further detail) and the mass profile for
the gas corresponds to $p$ = 2 if we invoke an outer boundary at $r_T
\approx r_s$.  A galactic disk with $R_C$ = 24 kpc would require an
initial rotation rate $\Omega \sim 10^{-16}$ sec$^{-1}$ and would
result in a disk mass $\sim 10^{11} M_\odot$.  This value of $\Omega$
is comparable to (but somewhat larger than) that expected for a spin
parameter $\lambda \approx 0.05$ (Bullock et al. 2001) for the same
halo.  This disk mass implies a total galactic mass of about $2.4
\times 10^{12} M_\odot$ (since only 1/4 of the gas mass collapses and
the ratio of baryons to total matter is about 1:6).\footnote{For
completeness we note that the baryon to dark matter ratio in galaxies
is lower than that of the universe as a whole by a factor of 3 -- 5. 
This difference can be explained by internal galaxy processes, e.g.,
long gas cooling times coupled with strong feedback.}  Figure
\ref{fig:profiles} also shows an exponential surface density profile
for comparison, where the scale length $\rexp$ = 5 kpc and the
exponential disk has the same total mass.  Compared to the exponential
disk, the surface density profiles resulting from these spherical
collapse solutions increase more steeply toward the center and
truncate sharply at the edge ($R_C$).  Nonetheless, over the
intermediate range of radii shown here, the surface density profiles
are roughly similar. Notice also that these profiles are similar to
those obtained in previous work based on conservation of angular
momentum only (see Fig. 1 of Dalcanton et al. 1997) and by detailed
SPH simulations of disk formation including many additional physical
processes (compare with Figs. 9, 10, and 16 of Kaufmann et al. 2006,
and with Figs. 10 and 11 of van den Bosch 2001).

Although we generalize this basic calculation of the surface density
in subsequent sections, the general features shown in Figure
\ref{fig:profiles} are robust, namely a power-law profile with a
relatively sharp truncation at the outer disk edge. This outer edge is
set by conservation of angular momentum. For a given starting state,
the gas has a maximum specific angular momentum $j_{max}$.  Material
with angular momentum $j = j_{max}$ will fall to a well-defined radius
(that of the disk edge) and material with $j < j_{max}$ will fall to
smaller radii. The surface density profiles tend to diverge in the
center $\varpi \to 0$. The centers of actual galaxies are complicated
by the presence of bulges (Binney \& Merrifield 1998) and central
black holes (Gebhardt et al. 2000). Further, this behavior can be
controlled by including a core radius in the initial density profile
(\S 4).

\begin{figure}
\figurenum{3}
{\centerline{\epsscale{1.25} \plottwo{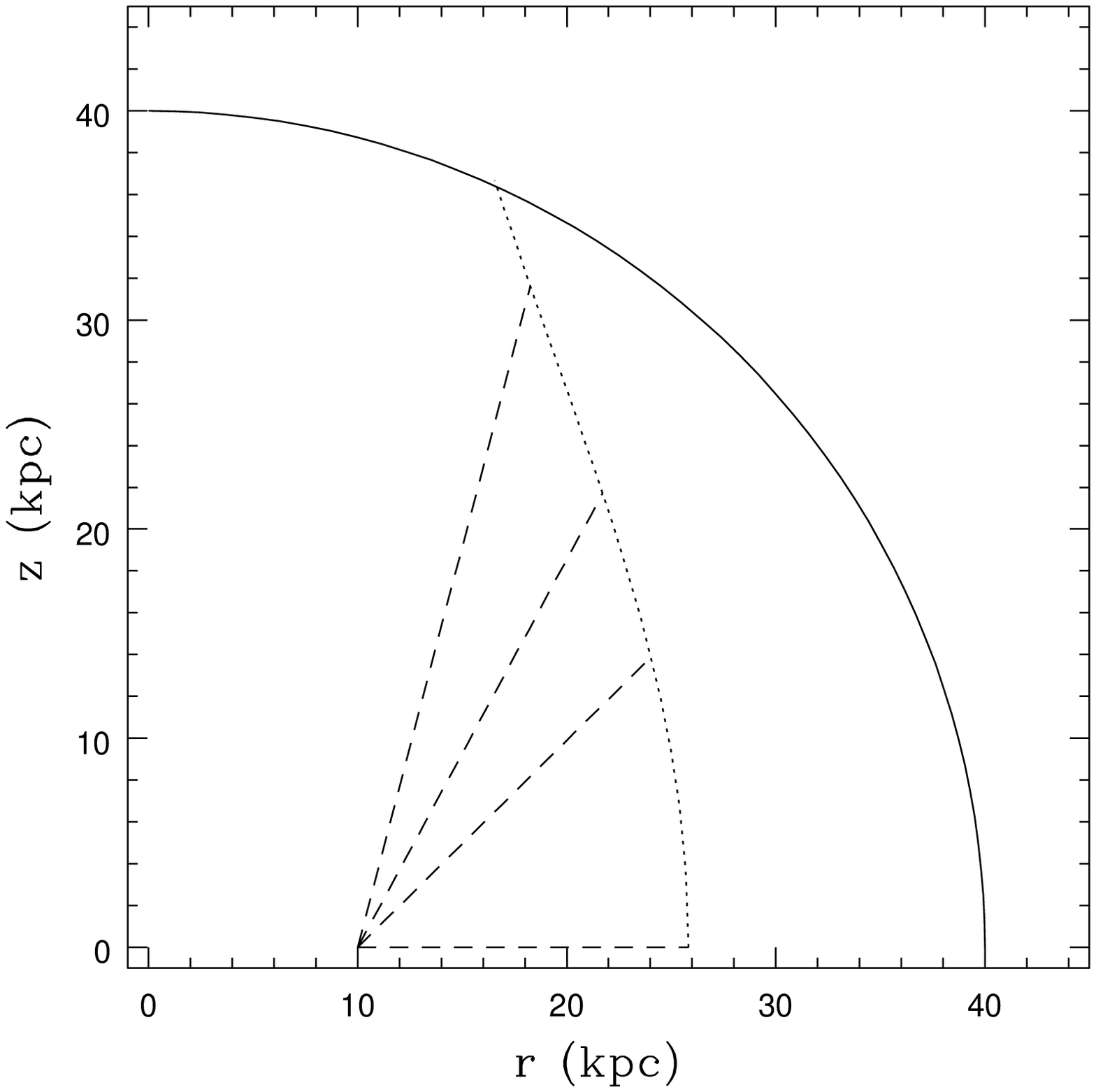} {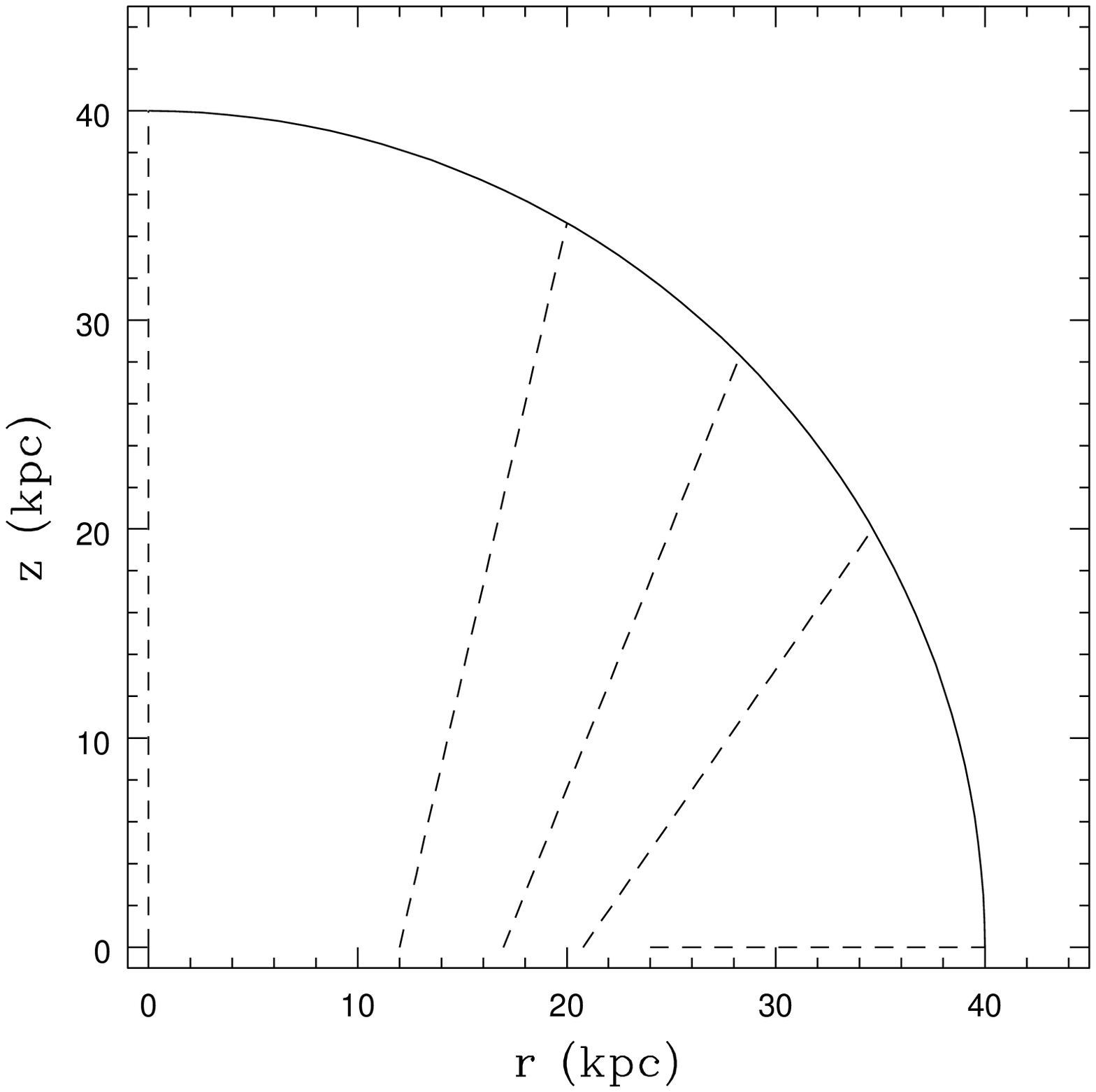} }}
\figcaption{Diagram showing the relationship between starting
locations for the orbits and the final locations. The left panel shows
the starting positions that lead to gas parcels falling to a given
radial location within the disk; the dotted curve shows the locus of
starting points for which the orbits land at the particular location
$\varpi$ = 10 kpc. The solid curve shows the outer boundary of the
collapsing region (taken here to be 40 kpc). The right panel shows the
final radial positions in the disk for starting points on a given
spherical surface, taken here to be the outer boundary at 40 kpc. In
both panels, the dashed lines are schematic and connect the starting
and ending locations.}
\label{fig:diagram} 
\end{figure}

In order to understand the infall picture in greater detail, it is
useful to see where gas parcels that start at various initial
locations end up within the disk.  Figure \ref{fig:diagram}
illustrates the general behavior (using the $p$ = 2 mass profile).
The surface density at a given radial location within the disk,
denoted by the coordinate $\varpi$, is fed by a range of initial radii
$r$ within the starting (spherical) state.  As the initial radius
increases, the initial polar angle $\theta$ decreases along a
well-defined locus of points, as shown in the left panel of Figure
\ref{fig:diagram}.  For a given radial location $\varpi$ in the disk,
there exists a minimum polar angle that contributes to the surface
density; for starting polar angles constrained by the relation
$\sin\theta < \varpi/R_C$, all the of material falls to smaller radii
$\varpi' < \varpi$.  One can also ask where the material at a given
starting radius ends up.  This result is shown by the right panel of
Figure \ref{fig:diagram}, where gas parcels along the outer boundary
of the collapsing region are mapped to their landing points on the
galactic disk. The pole ($\theta = 0$) is always mapped to the
galactic center and the material along the equator falls to the outer
disk edge (in this geometry, the material along the equator has the
highest specific angular momentum and defines the outer disk edge).

\subsection{General spherical starting states with constant rotation rate}  

The treatment developed here can be generalized to include any
power-law mass profile, i.e.,
\be 
M(r) \propto r^p \qquad {\rm and} \qquad r(M) \propto M^{1/p} \, . 
\ee 
We define the dimensionless variable $x \equiv (M/M_s) (\omega/\xi)^{p/2}$, 
and the surface density profile can be written in the form 
\be 
\sigma (\varpi) = {M_T \over 2 \pi \varpi^{2-p/2} R_C^{p/2} } \, 
\int_1^{x_f} {dx \over x^{4/p}} \bigl[ 1 - x^{-{4/p} } \bigr]^{-1/2} \, , 
\label{eq:siggeneral} 
\ee
where 
\be
R_C = \rot (M_T/M_s)^{2/p} \, , 
\ee  
so that $x_f = (R_C/\varpi)^{p/2}$.  

As a consistency check, we can show that the total mass $M_D$ locked
up in the surface density profile is equal to the original mass $M_T$
that falls onto the disk. To show this, we integrate the general form
of the surface density (eq. [\ref{eq:siggeneral}]) over the surface of
the disk to obtain
\be
M_D = \int_0^{R_C} 2 \pi \varpi d\varpi 
{M_s \over 2 \pi \varpi^{2-p/2} \rot^{p/2} } \, 
\int_1^{x_f} {dx \over x^{4/p}} \bigl[ 1 - x^{-{4/p} } \bigr]^{-1/2} \, . 
\ee
The outer edge of the disk is given by $R_C$ = $\rot (M_T/M_s)^{2/p}$ 
and the upper limit of integration in the first integral can be 
written as $x_f$ = $(R_C/\varpi)^{p/2}$. If we change the integration 
variable from $\varpi$ to $z \equiv \varpi/R_C$, the integral becomes 
\be
M_D = M_T \int_0^1 {dz \over z^{1-p/2}} \int_0^{(1/z)^{p/2}} 
{dx \over x^{4/p}} \bigl[ 1 - x^{-{4/p} } \bigr]^{-1/2} \, . 
\ee 
Next, we change the order of integration and evaluate: 
\be
{M_D \over M_T} = \int_1^\infty {dx \over x^{4/p}} 
\bigl[ 1 - x^{-{4/p} } \bigr]^{-1/2} \int_0^{(1/x)^{2/p}} 
{dz \over z^{1-p/2}} 
= {2 \over p} \int_1^\infty {dx \over x} 
x^{-4/p} \bigl[ 1 - x^{-{4/p} } \bigr]^{-1/2} \, 
= 1 \, . 
\ee 
This argument shows that mass is conserved for any mass profile, 
as characterized by the index $p$, i.e., the resulting surface 
density profile has the same total mass as the starting state.  

The general result of this analysis is that a single spherical
collapse naturally produces a power-law surface density profile with a
sharp truncation at the outer edge. We can be more precise about the
sharpness of the edge: For any value of $p$, we obtain a integral $I$
(see eq. [\ref{eq:siggeneral}]) that is a function of $x_f$ and hence
a function of the radial disk coordinate $\varpi$.  The variation of
the surface density profile can be characterized by the index $Q$
defined by $Q \equiv (\varpi/I) (d I / d \varpi)$.  It is
straightforward to show that $Q \to \infty$ in the limit that $\varpi
\to R_C$ for all indices $p > 0$.

\subsection{Starting states with constant rotation speed} 

Many numerical simulations of structure formation suggest that the
angular momentum distribution of the gas, before collapse, takes the
form $j (r) \propto r$ (Kaufmann et al. 2006; Bullock et al. 2001).
This angular momentum profile is equivalent to a rotation rate 
with radial dependence $\Omega \propto r^{-1}$, rather than the
simplest case of $\Omega$ = {\sl constant} considered above. We can
generalize the previous treatment as follows.  The rotation rate of
the starting gas can be written in the form
\be 
\Omega (r) = \Omega_s {r_s \over r} \, = \Omega_s {1 \over \xi} \, , 
\ee
where we have scaled the angular velocity to its value at $r = r_s$.
For this rotation profile, the angular momentum parameter $\qam_0$ (see
eq. [\ref{eq:qzero}]) takes the form 
\be 
\qam_0 = {r_s^2 \Omega_s^2 \over 2 \Psi_0} 
\xi_\infty^2 = \omega^2 \xi_\infty^2 
\qquad {\rm where} \qquad 
\omega = {r_s \Omega_s \over \sqrt{2 \Psi_0}} \, . 
\label{eq:soneomega} 
\ee
For a given mass profile of the form $M \propto r^p$, 
the integral expression that determines the surface density 
profile takes the form 
\be 
\sigma (\varpi) = {M_s \over 2 \pi \varpi^{2-p} \rot^{p} } 
\int_1^{x_f} {dx \over x^{2/p} } \bigl[ 1 - x^{-2/p} \bigr]^{-1/2} \, , 
\ee 
where the integration variable is given by $x = (M / M_s)
(\rot/\varpi)^p$.  For these models, the centrifugal radius, which
sets the location of the outer disk edge, is given by
\be 
R_C = \rot (M_T/M_s)^{1/p} \, , 
\label{eq:sonerc} 
\ee 
and hence $x_f$ = $(R_C/\varpi)^p$. 

For this choice of initial rotation profile, the surface density
integrals can be evaluated in terms of elementary functions for
starting mass profiles with index $p$ = 1, 2, and 3. For the starting
density profile of an isothermal sphere, $p$ = 1, we obtain
\be
\sigma (\varpi) = {M_T \over 2 \pi \varpi R_C} \cos^{-1} (\varpi/R_C) \, , 
\label{eq:sonepone} 
\ee 
which is the same as that obtained earlier for the case of the
intermediate $p$ = 2 mass profile and a constant rotation rate. With
this rotation profile $\Omega \sim r^{-1}$ and $p$ = 2, the surface
density takes the form
\be 
\sigma (\varpi) = {M_T \over \pi R_C^2} \cosh^{-1} (R_C/\varpi) \, . 
\label{eq:soneptwo} 
\ee 
For a uniform density starting state, $p$ = 3, the surface density
becomes
\be 
\sigma (\varpi) = {3 M_T \over 2 \pi R_C^2} \, 
\bigl( 1 - \varpi^2 / R_C^2 \bigr)^{1/2} \, . 
\label{eq:sonepthree} 
\ee 
Note that equations (\ref{eq:sonepone} -- \ref{eq:sonepthree})
describe how the surface density profiles (for constant rotation
speed) depend on the underlying parameters (where the definitions of
eqs. [\ref{eq:soneomega}] and [\ref{eq:sonerc}] complete the
specification).

\begin{figure}
\figurenum{4}
{\centerline{\epsscale{0.90} \plotone{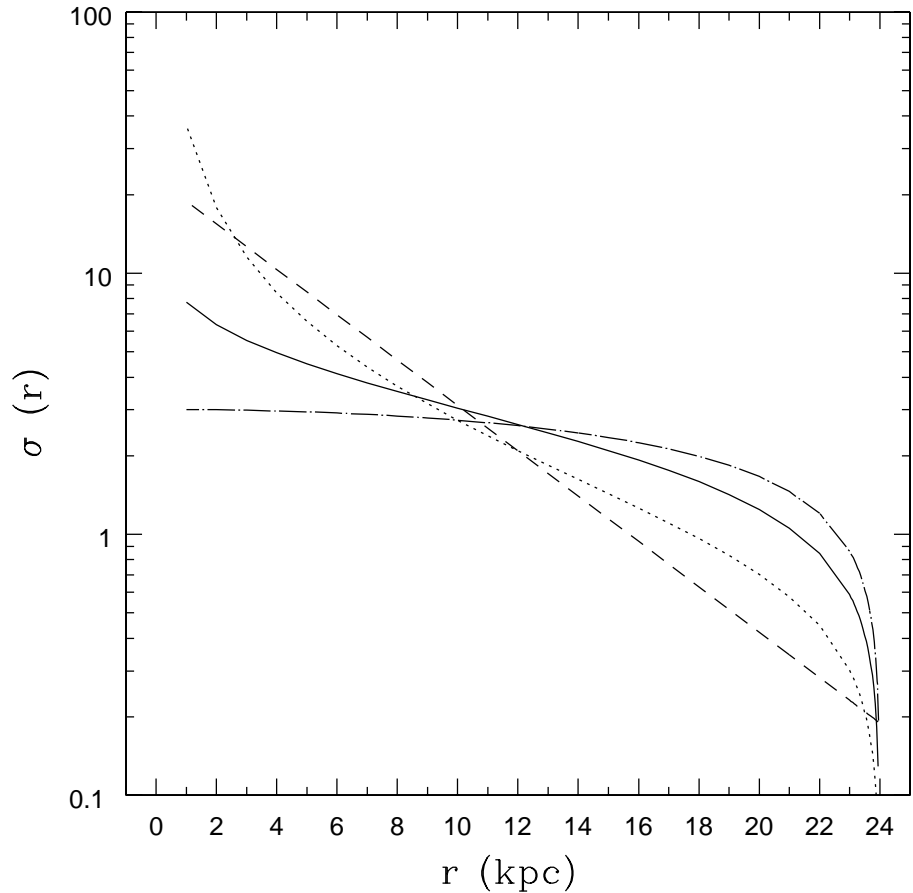} }}
\figcaption{Surface density profiles resulting from spherical collapse
with starting rotation rate $\Omega \propto r^{-1}$.  The resulting
surface density is shown for three values of the index $p$ of the
starting mass profile: $p$ = 1 (dotted curve), $p$ = 2 (solid curve),
and $p$ = 3 (dot-dashed curve).  All three profiles have the same
total mass and the same value of the centrifugal barrier ($R_C$ = 24
kpc), which determines the outer disk edge.  The dashed curve shows an
exponential profile for comparison, where the exponential scale length
is $\rexp$ = 5 kpc and the total disk mass is the same as the other
profiles. The scale on the vertical axis is arbitrary.} 
\label{fig:profiles2} 
\end{figure}

Figure \ref{fig:profiles2} shows the surface density profiles for the
three mass profiles $p$ = 1, 2, and 3, for the case of constant
starting rotation speed ($\Omega \sim r^{-1}$). These profiles are
less steep than those with constant rotation rate ($\Omega$ = {\sl
constant}; see Fig. \ref{fig:profiles}). In this case, the $p$ = 1
starting state leads to a power-law surface density profile $\sigma
\sim \varpi^{-1}$ and the $p$ = 3 starting states leads to $\sigma
\sim \varpi^0$ (where both power-laws are truncated at $R_C$ by an
edge function). The intermediate mass profile with $p$ = 2 produces a
surface density profile of the approximate form $\sigma \sim \log
(R_C/\varpi)$ away from the disk edge (i.e., for $\varpi \ll R_C$).
For $r_T \approx r_s$, a value of the constant rotation speed $v_s =
r_s \Omega_s \approx 180 - 200$ km/s is required to produce $R_C$ = 24
kpc as shown here.

\subsection{General spherical starting states} 

This subsection considers a general spherical starting state in which
the mass profile has the form $M \propto r^p$ (with $p$ in the range
$1 \le p \le 3$, but otherwise arbitrary) and the rotation profile has
the form $\Omega \propto r^{-s}$ (with $s$ in the range $0 \le s \le
1$). The centrifugal radius takes the form
\be 
R_C = \rot \Bigl( {M_T \over M_s} \Bigr)^{1/ \alpha} 
\qquad {\rm where} \qquad \alpha = {p \over 2 - s} \, . 
\ee 
The index $\alpha$ plays an important role in determining the 
functional form of the surface density, which is given by 
\be 
\sigma(\varpi) = {M_T \over 2 \pi \varpi^{2 - \alpha} R_C^\alpha }
\int_1^{x_f} {dx \over x^{2\alpha} } \bigl[ 1 - x^{-2\alpha}
\bigr]^{-1/2} \, .
\ee 
The effective power-law index $q$ of the surface density profile (far from 
the disk edge at $R_C$) is given by 
\be 
q \equiv 2 - \alpha = 2 - {p \over 2 - s} \, . 
\label{eq:qgeneral} 
\ee 
This interpretation (that $q$ is the power-law index of the profile) 
breaks down when $q \to 0$, i.e., the portion of parameter space 
defined by $q < 0$ results in qualitatively different surface density 
profiles (ones that cannot be described in terms of power-laws). In 
general, these surface density profiles are shallower than power-laws. 

\subsection{\bf Hydrostatic initial states} 

This subsection considers the scenario in which gas enters the central
region of a dark matter halo and reaches hydrostatic equilibrium
before cooling and falling inward to the disk. In this approximation,
the gravitational potential is provided by the dark matter halo alone.
The resulting (pre-collapse) density profile depends on the equation
of state for the gas. We first consider the idealized case of an
isothermal gas with a uniform rotation rate; we then consider the more
realistic case of a general polytropic equation of state with a
constant velocity ($\Omega \propto 1/r$) initial condition. 

In the limit of an isothermal equation of state, the starting
hydrostatic density profile takes the form
\be 
\rho (r) = \rho_\infty \exp \big[ 
{(\Psi_0 / a^2) \over 1 + \xi} \bigr] \, = 
\rho_0 \exp \Bigl[ - \Bigl( {\Psi_0 \over a^2} \Bigr) 
{\xi \over 1 + \xi} \Bigr] \, , 
\label{eq:isodense} 
\ee 
where $a$ is the isothermal sound speed of the gas. After cooling, the
gas falls through the potential of the dark matter halo and collects
into a disk structure. In this case, the integral that determines the
disk surface density is more naturally written as a radial integral, 
rather than a mass integral. We assume that the starting (pre-collapse) 
density profile follows the form of equation (\ref{eq:isodense}) out 
to an outer boundary radius $r_T$. We then work in terms of the 
variable defined by $y \equiv (r/r_s) (\rot/\varpi)^{1/2}$, and 
define the centrifugal radius via
\be
R_C = \rot (r_T/r_s)^2 \, . 
\ee 
Note that we have assumed a uniform rotation rate in deriving this 
form for $R_C$. The integral that determines the disk surface density 
then takes the form 
\be
\sigma (\varpi) = {2 r_T^3 \rho_0 \over R_C (R_C \varpi)^{1/2} } 
\int_1^{y_f} {dy \over (y^4 - 1)^{1/2} } \exp \Bigl[ - \Bigr( 
{\Psi_0 \over a^2} \Bigr) {\lambda y \over 1 + \lambda y} \Bigr] \, ,  
\label{eq:isoint} 
\ee 
where $\lambda \equiv (\varpi/R_C)^{1/2} (r_T/r_s)$. 

\begin{figure}
\figurenum{5}
{\centerline{\epsscale{0.90} \plotone{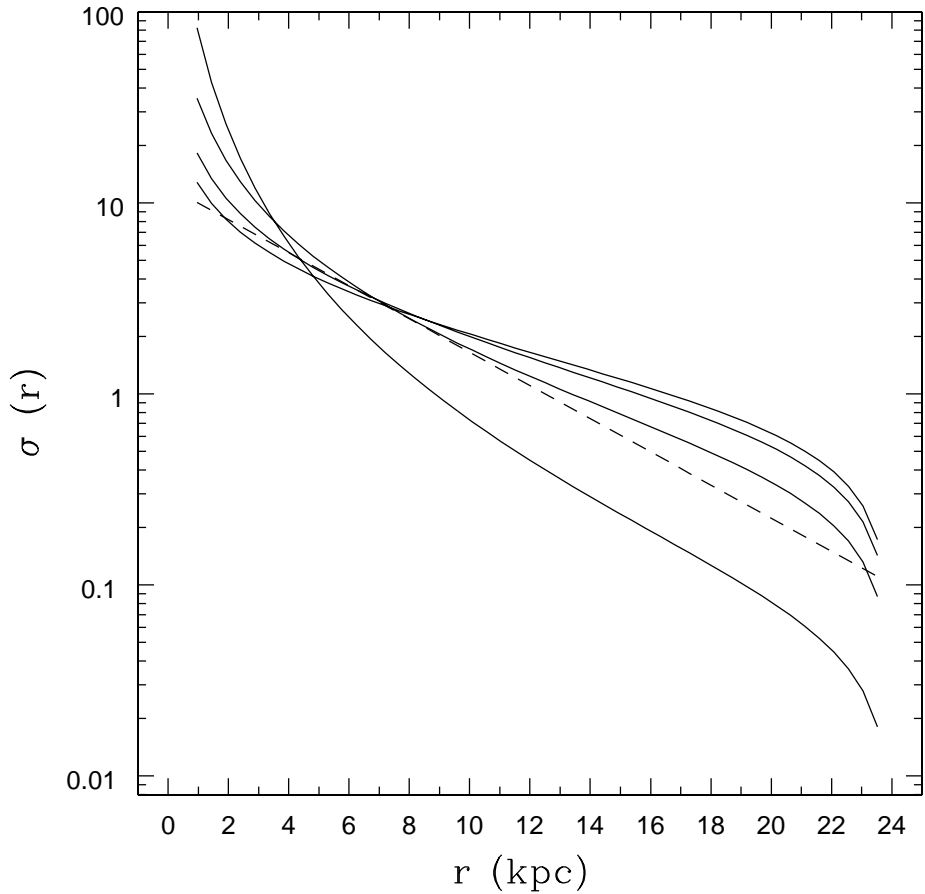} }} 
\figcaption{Surface density profiles resulting from spherical 
collapse starting from quasi-static equilibrium of an isothermal gas
in a dark matter halo. The four curves show the surface density
profiles for varying values of the ratio $\Psi_0/a^2$ = 2, 4, 8, and
16, where $a$ is the isothermal sound speed of the gas (before it
cools and collapses). The surface density distribution becomes steeper
as the ratio $\Psi_0/a^2$ increases. Cold gas (with sound speed $a$
small compared to the depth of the gravitational potential well)
results in a steep surface density profile because mass is
concentrated at small radii in the hydrostatic (pre-collapse) state.
The dashed curve shows an exponential profile for reference, where the
scale length $\rexp$ = 5 kpc. The scale on the vertical axis is arbitrary. } 
\label{fig:isodisk} 
\end{figure}

Figure \ref{fig:isodisk} shows the surface density distributions
resulting from this type of collapse, where the integral in equation
(\ref{eq:isoint}) has been evaluated numerically. For the cases shown
here, the initial radial extent of the pre-collapse gas was taken to
be $r_T$ = $r_s$, the rotation rate was chosen so that the centrifugal
radius $R_C$ = 24 kpc (i.e., $\Omega \sim 10^{-16}$ rad$^{-1}$ for the
Galactic parameters used previously), and all profiles have the same
total mass. The four curves shown in Figure \ref{fig:isodisk}
correspond to varying values of the ratio $\Psi_0/a^2$ = 2, 4, 8, and
16 (from top to bottom in the figure). For moderate values of the
ratio $\Psi_0/a^2$, the surface density distributions have the same
general form as those obtained previously (compare with
Figs. \ref{fig:profiles} and \ref{fig:profiles2}).

This approach can be generalized to include any polytropic equation of
state for the initially hydrostatic gas and any rotation profile
$\Omega(r)$.  We start by writing the equation of state in the form
\be 
P = \kappa \rho^\gamma \, = \kappa \rho^{1 + 1/n} \, , 
\ee 
where $n$ is the polytropic index (e.g., Chandrasekhar 1939). 
The density distribution for hydrostatic equilibrium, in the 
limit where the dark matter halo dominates the gravitational 
potential, then takes the form 
\be
\rho = { \rho_0 \over (1 + \xi)^n } \, , 
\label{eq:polydense} 
\ee 
where the central density is specified through the relation 
\be
\rho_0 = \bigl[ \Psi_0 n / (n + 1) \kappa \bigr]^n \, . 
\ee 
In the limit $n \to \infty$, the equation of state approaches an
isothermal form; in the limit $n \to 3/2$, $\gamma \to 5/3$, and 
the equation of state becomes adiabatic (for a monatomic gas).  
The polytropic index is thus confined to the range $3/2 \le n <
\infty$. We can see immediately that the value $n$ = 3 ($\gamma$ =
4/3) corresponds to a critical value: If $n > 3$, then the mass
integral converges, whereas the mass integral diverges for $n \le 3$.

\begin{figure}
\figurenum{6}
{\centerline{\epsscale{0.90} \plotone{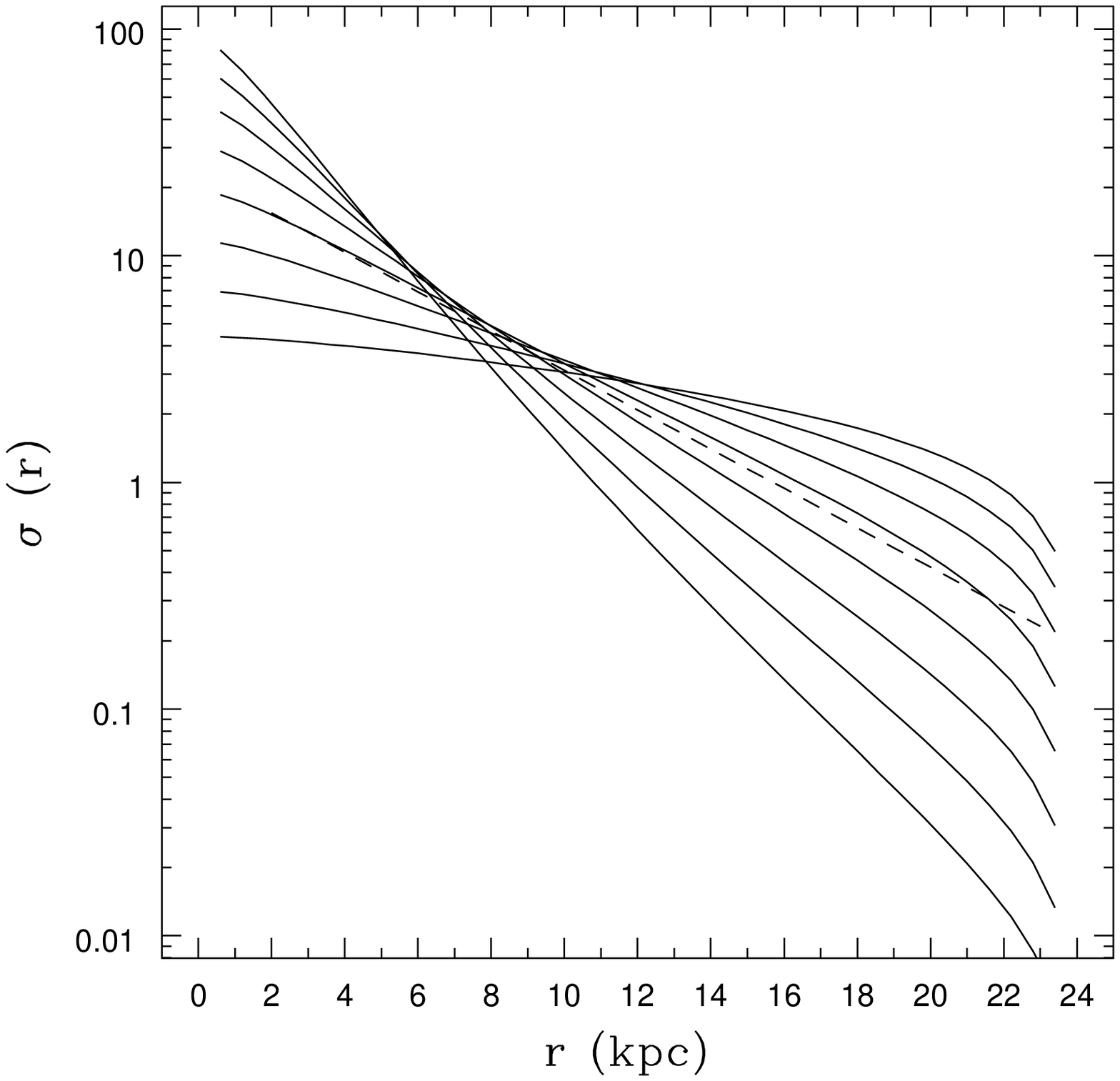} }} 
\figcaption{Surface density profiles resulting from spherical collapse
starting from quasi-static equilibrium in a dark matter halo where the
gas has a polytropic equation of state and the starting rotation
profile $\Omega (r) \propto 1/r$. The profiles shown in the figure
correspond to polytropic indices $n$ = 2 -- 16 (from top to
bottom). All of the profiles are normalized to have the same mass.  As
the polytropic index $n$ increases, the equation of state becomes
closer to isothermal, more gas is concentrated at small radii in the
pre-collapse state, and the disk surface density becomes steeper.  The
dashed curve shows an exponential profile with scale length $\rexp$ =
5 kpc. The slope of the surface density produced from the collapse of
the $n=8$ polytrope is close to this reference profile.  The scale on
the vertical axis is arbitrary. }
\label{fig:polytrope} 
\end{figure}

Following the same formulation used above for the isothermal case, the
disk surface density resulting from the collapse of a polytropic
sphere with rotation profile $\Omega \propto 1/r$ takes the form
\be 
\sigma(\varpi) = {2 \rho_0 r_s^3 \varpi \over \rot^3} 
\int_1^{y_f} {y \, dy \over (y^2 - 1)^{1/2} } 
{1 \over (1 + \lambda y)^n } \, , 
\ee
where $y_f$ = $(r_T/r_s) (\rot/\varpi)$ and $\lambda$ =
$(\varpi/\rot)$.  As a consistency check, we can integrate this
expression over the surface of the disk to find the total disk mass,
and thereby obtain the expected result
\be
M_D = \int_0^{R_C} 2 \pi \varpi d\varpi \sigma (\varpi) = 
4 \pi \rho_0 r_s^3 \int_0^{r_T/r_s} {\xi^2 d\xi \over 
(1 + \xi)^n} \, = \, M(r_T) \, . 
\ee 

For any polytropic index $n$, the resulting surface density
distributions display the same general form -- with radial dependence
somewhere between a truncated power-law and an exponential. In these
disks, the outer radius is determined by the centrifugal barrier,
which in turn is determined by the material with the highest specific
angular momentum, and that material comes from the equatorial region
at the outer boundary $r_T$ of the collapsing region. A collection of
surface density profiles of this type is shown in Figure
\ref{fig:polytrope} for polytropic indices $n$ = 2 -- 16. For the
profiles shown here, the central density $\rho_0$ is assumed to be
constant and the profiles are normalized to have constant total
mass. For these parameter choices, the surface density profiles become
steeper as the polytropic index $n$ increases. The resulting curves
are close to straight lines on the plot of $\log\sigma$ versus
$\varpi$, where the effective scale radius of the exponential
decreases with increasing polytropic index $n$. Figure
\ref{fig:polytrope} also shows an exponential profile with scale
length $\rexp$ = 5 kpc (dashed curve). The surface density profile
produced by the collapse of the $n$ = 8 polytrope has nearly the same
slope. For comparison, models of galaxy groups and clusters have been
used to estimate the equation of state in astrophysical systems
(Ascasibar et al. 2003) and find a fit of the form $P \propto
\rho^{1.18}$, which corresponds to a polytropic index $n \approx 5.6$.

For completeness we note that for all polytropic indices $n > 3$, the
total mass of the pre-collapse gaseous sphere is finite and disk
solutions exist for the limit $y_f \to \infty$. In this case, the
resulting surface density profiles approach power-law forms such that
$\sigma \sim \varpi^{-(n-1)}$ at large radii. However, the convergence
to this power-law form is painfully slow.

\subsection{Section summary}  

Actual galactic disks differ from the idealized disks found in this
section in several ways. As outlined below, the single, spherical
collapse picture must be generalized to allow for more complicated
starting conditions (e.g., holes and filaments; see \S 4) and multiple
disk accretion events (e.g., due to merging of subhalos, which supply
additional gas; see \S 5).  This work makes additional assumptions
that can be studied in future work. We have not taken into account
adiabatic contraction of the dark matter halo.  This effect will make
the gravitational potential well deeper and could be taken into
account by changing the values of $\Psi_0$ in the orbit solutions. A
deeper potential well leads to greater infall velocities. Since the
perpendicular component of the velocity is dissipated at the disk
surface, the energy loss is greater for deeper potentials. For a given
angular momentum, a deeper potential leads to a smaller centrifugal
radius for a given orbit; notice, however, that the angular momentum
of later orbits may be larger as well (e.g., Elmegreen et al. 2005)
and lead to a larger centrifugal radius.  This current treatment also
makes the approximation of zero energy orbits and considers orbital
solutions in the inner part of the potential. The latter approximation
is not severe, since the orbits must be nearly radial in the outer
part of the potential in order to reach the disk. Similarly, if the
initial energy is small but nonzero, the orbit will be largely
unchanged (where the basic constraint is $\epsilon = |E|/\Psi_0 \ll 1$).

In addition, the calculation thus far determines the surface density
profile resulting from the initial infall, i.e., the incoming gas
parcels are assumed to stay at the radial locations where they fall.
Over time, the disk will dissipate energy, transfer angular momentum,
and spread out. Notice that a moderate amount of disk spreading
(angular momentum transfer) will make the calculated surface density
profiles look more like the exponential reference profile over the
range of intermediate radii (roughly 2 -- 20 kpc for the parameters
used here).  Notice also that real disk galaxies depart from an
exponential law both on the inside and on the outside. The presence of
galactic bulges causes the brightness profiles of observed disk
galaxies to lie above that expected for an exponential disk at small
radii, 1--2 kpc (see Binney \& Merrifield 1998 for a detailed
discussion of disk/bulge decomposition). On the outside, some disk
galaxies show evidence for a cutoff radius $R_{max}$ that could be
identified with $R_C$ in this theoretical treatment; typical values
are $R_{max} \approx 20 - 25$ kpc (van der Kruit \& Searle 1981;
Wainscot et al. 1988; Barteldrees \& Dettmar 1994; Binney \&
Merrifield 1998). 

\section{GENERALIZED COLLAPSE SOLUTIONS} 

\subsection{Partial radial coverage -- shells and holes}

This subsection considers configurations for which the initial
(pre-collapse) gaseous sphere has holes, i.e., an evacuated central
region. Suppose that the central portion of the sphere, with radius 
$r < r_0$, has no gas. The corresponding surface density profile is
that of the original sphere (with no central hole) with the
contribution from a smaller sphere ($0 \le r \le r_0$) removed. If the
original sphere has radius $r_T$ and mass $M_T = M(r_T)$, and the hole
region has radius $r_0$ and mass $M_0 = M(r_0)$, the combined surface
density profile is given by
\be 
\sigma (\varpi) = \sigma(\varpi; M_T) - \sigma(\varpi; M_0) \, , 
\ee
where the surface density profiles on the right hand side of the 
equation are those obtained via the spherical collapse calculation 
of the previous section. 

\begin{figure}
\figurenum{7}
{\centerline{\epsscale{0.90} \plotone{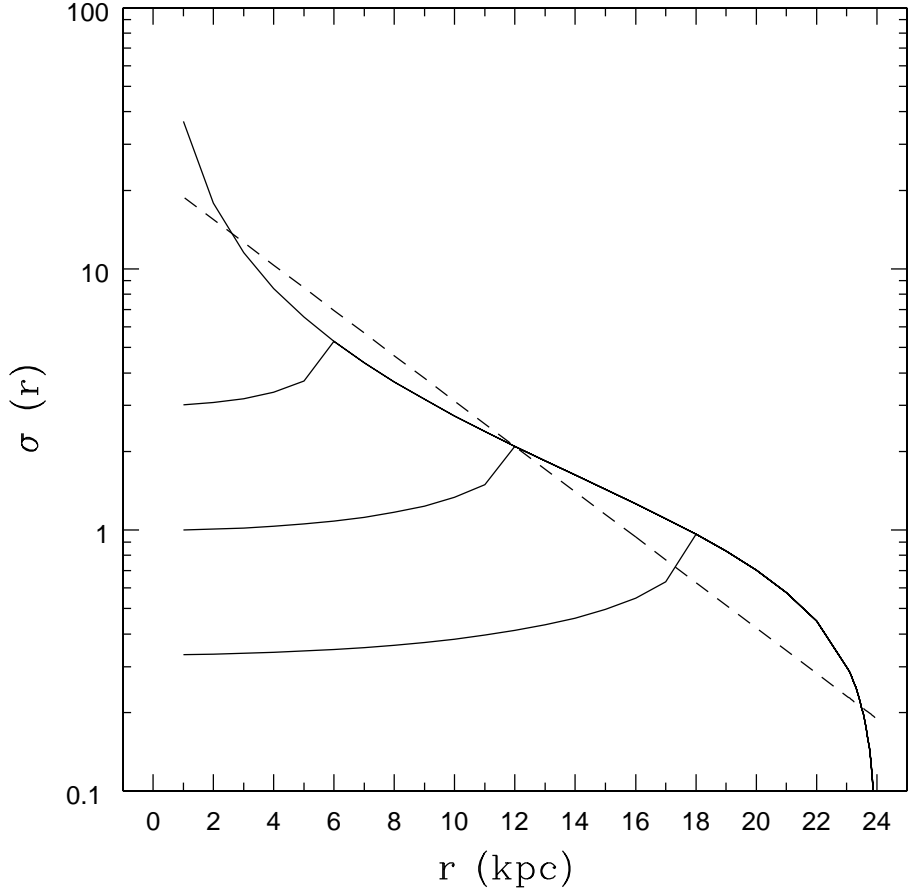} }}
\figcaption{The effects of removing the central regions from the
starting mass profile on the resulting surface density profiles. The
solid curves show the result of a generalized collapse with $R_C$ = 24
kpc and with varying amounts of mass removed from the central regions.
The upper curve shows the baseline result for a complete gaseous
sphere.  The other three solid curves show the surface density
profiles when a central fraction $f$ of the mass of the starting
sphere is removed, where $f$ = 0.25, 0.50, and 0.75 (from top to
bottom).  The dashed curve shows an exponential surface density
profile with the same mass as that of the complete spherical starting
state and with scale length $\rexp$ = 5 kpc. The scale on the vertical 
axis is arbitrary. }
\label{fig:radholes} 
\end{figure}

One example of collapse with a centrally evacuated region is shown in
Figure \ref{fig:radholes}. Here we consider the intermediate mass
profile with $p$ = 2 ($M \sim r^2$) with constant rotation rate, and
remove successively larger holes from the center of the starting
state.  The four curves shown in the figure correspond to the full
sphere (top curve) and increasing fractions of the mass evacuated
(25\% to 75\%).  For comparison, the dashed curve shows an exponential
surface density profile with the same mass as the surface density
resulting from the collapse of the full sphere. As before, the
rotation rate is chosen so that $R_C$ = 24 kpc.  

\subsection{Partial angular coverage}

We can also use this formalism to determine the surface density
profiles resulting from pre-collapse states with partial angular
coverage. In general, the starting mass distribution can be restricted
in both azimuthal angle $\varphi$ and the polar angle $\theta$.  In
this problem, the starting azimuthal angle determines the azimuthal
angle where a given gas parcel will join the galactic disk, but it
otherwise plays no role (e.g., starting angular momentum depends on
$(r,\theta)$ but not $\varphi$). Furthermore, after gaseous material
has joined the disk, the differential rotation within the disk will
eventually make the disk axisymmetric, thereby erasing any dependence
on $\varphi$. Thus, if we consider starting states that are restricted
in azimuthal angle $\varphi$ (spheres with longitudinal slices
removed), the net result is to reduce the total mass in the disk by a
constant factor, but otherwise leave the surface density unchanged. We
thus consider the case of restricting the starting state in polar
angle $\theta$.

\begin{figure}
\figurenum{8}
{\centerline{\epsscale{0.90} \plotone{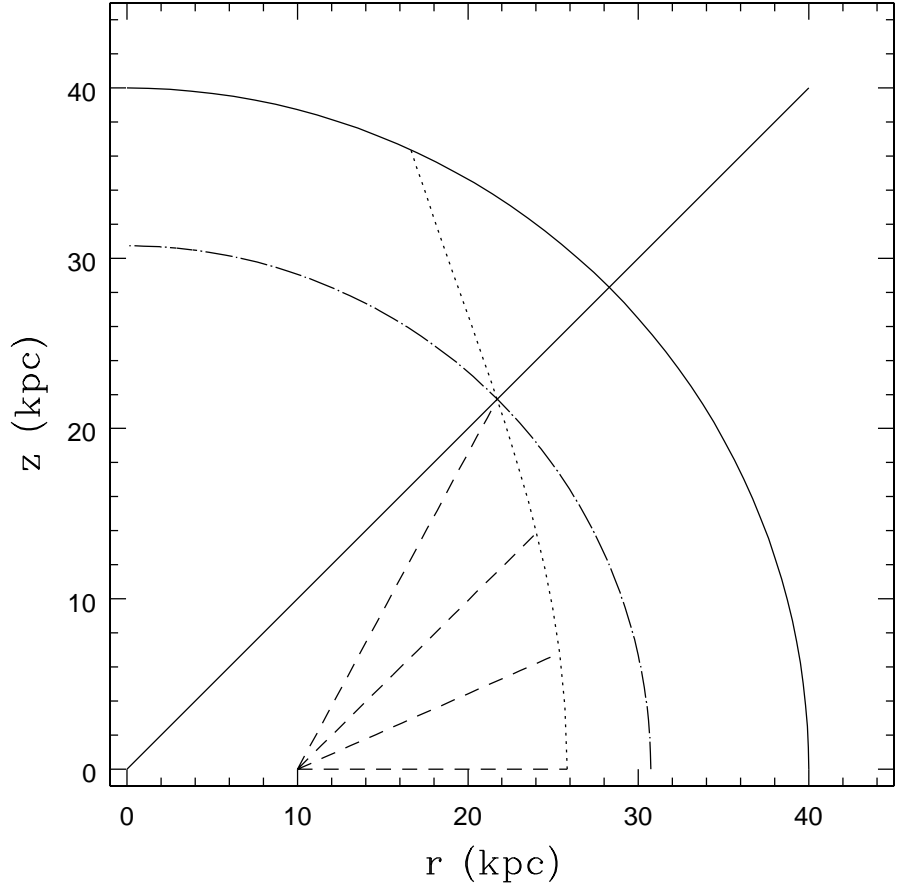} }} 
\figcaption{Diagram showing the geometry for infall with a polar cone
removed. The outer solid circular curves denotes the outer boundary of
the collapsing region. The solid ray, drawn here at $\theta_M$ =
$\pi/4$, delineates the evacuated region ($\theta < \theta_M$). The
dotted curve shows the locus of starting points that fall to a
particular radius within the disk (taken here to be $\varpi$ = 10
kpc). The dashed lines connect the starting locations with the final
positions. For a given radius in the disk, the surface density is the
same as that resulting from the collapse of a smaller sphere, one with
radius given by the intersection of the ray $\theta = \theta_M$ with
the locus of starting points; the dot-dashed circular curve shows this
effective boundary for the radial location $\varpi$ = 10 kpc in the
disk. } 
\label{fig:conediagram} 
\end{figure}

Suppose we remove the polar cones from the starting state so that the
polar angle is restricted to the range $\theta_M \le \theta \le \pi -
\theta_M$. The surface density contribution at a given radial location
$\varpi$ in the disk is the same as that produced by the collapse of
the inner portion of the sphere with mass $M(r = r_{\theta M})$, where
$r_{\theta M}$ is given by the intersection of the ray $\theta =
\theta_M$ and the locus of points for which material falls to radial
position $\varpi$, i.e., the locus given by
\be
r_s \sqrt{\qam_0} \sin \theta_0 = \varpi \, . 
\ee 
 
Consider the case of an initial mass profile with $p$ = 2 ($M\sim
r^2$) and constant rotation rate.  If the starting state were a
complete sphere it would have mass $M_T$ and outer (spherical)
boundary $r_T$. In this case, the polar cones are removed, where
$\theta_M$ defines the opening angle of the cones. At a given 
radius $\varpi$ in the disk, the surface density is the same as 
that for mass 
\be
M = M_T {\varpi \over R_C \sin \theta_M} \, . 
\ee 
The surface density profile is thus given by 
\be 
\sigma (\varpi) = {M_T \over 2 \pi R_C \varpi} \int_1^{x_f} 
{dx \over x} \bigl[ x^2 - 1 \bigr]^{-1/2} \, , 
\ee 
where the upper limit of the integration is given by 
\be 
x_f = {\rm min} \Bigl\{ {R_C \over \varpi} , 
{1 \over \sin \theta_M} \Bigr\} \, . 
\ee 

\begin{figure}
\figurenum{9}
{\centerline{\epsscale{0.90} \plotone{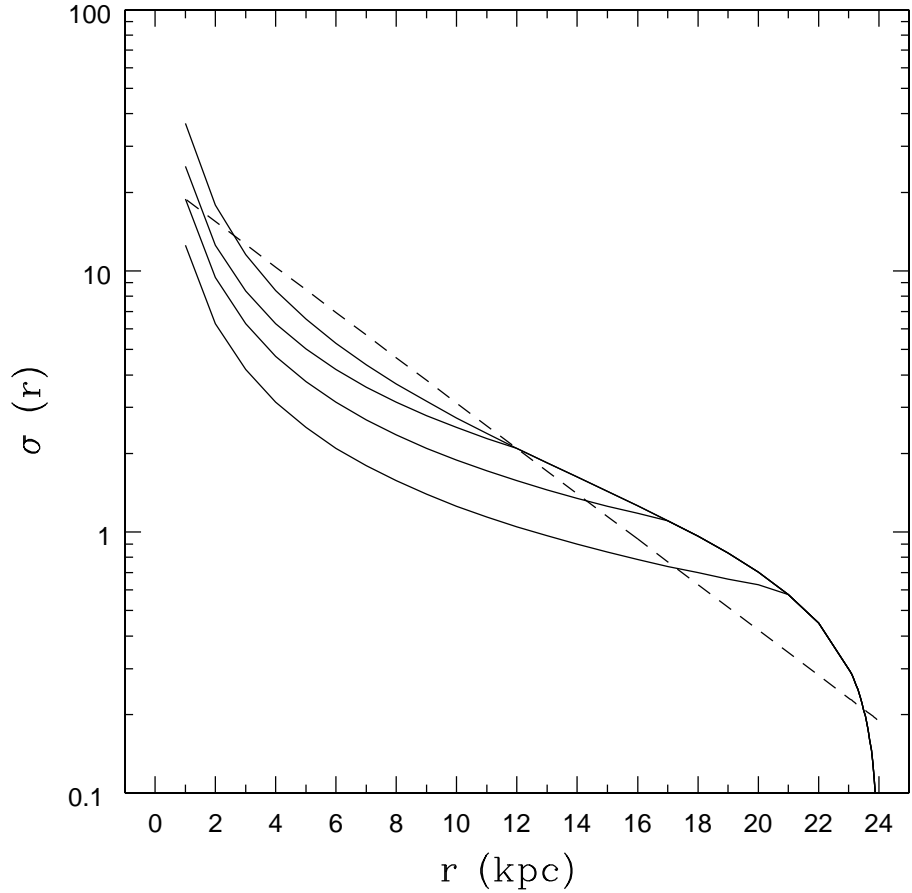} }}
\figcaption{The effects of removing the polar caps from the starting
mass profile on the resulting surface density profiles. The solid
curves show the result of an axisymmetric collapse with $R_C$ = 24
kpc.  The upper curve shows the result for a complete gaseous sphere
with a $p$ = 2 mass profile and constant rotation rate. The next three
solid curves show the surface density profiles for starting states
with polar caps removed, where the polar angle of removed material
$\theta_M$ = $\pi/6, \pi/4, \pi/3$. The dashed curve shows an
exponential surface density profile with the same mass as that of the
complete spherical starting state and with scale length $\rexp$ = 5
kpc. The scale on the vertical axis is arbitrary. }
\label{fig:polarholes} 
\end{figure}

Figure \ref{fig:polarholes} shows the resulting surface density
profiles for the intermediate starting mass profile with $M(r) \sim
r^2$ and constant rotation rate. The top curve shows the surface
density profile for the complete sphere. The next three curves
illustrate the effect of removing the polar cones from the starting
configuration, where the opening angle is taken to be $\theta_M$ =
$\pi/6$, $\pi/4$, and $\pi/3$. For these three cases, the fraction
$f_m$ of the total mass (within the same outer boundary) contained in
the initial configuration, and hence the resulting disk, is $f_m$ =
$\sqrt{3}/2$, $\sqrt{2}/2$, and $1/2$, respectively.

\subsection{The Limiting Case of Filaments} 

If we restrict the angular extent of a rotating sphere to its limiting
form, the result is a filament, which provides a reference state for
the (pre-collapse) initial conditions. Notice that a filament can be
defined in a variety of ways. This treatment starts with a spherical
initial condition, as considered above, and confines the geometry in
both polar angle $\theta$ and azimuthal angle $\varphi$.  In the
limiting case, the resulting filament has all of its mass along a
fixed ray defined by $\theta$ = {\sl constant} and $\varphi$ = {\sl
constant}. Further, this discussion is limited to cases where the
rotation rate $\Omega$ = {\sl constant}, so that the filament has some
angular momentum about the galactic center but it remains intact as it
rotates. These filaments are thus highly idealized, but provide a
useful benchmark for understanding the nature of galactic infall.
Keep in mind, however, that these filaments are not the large scale
structures (also called filaments) that connect galaxies to the cosmic
web.

We begin by considering an isothermal initial configuration, so that
$M(r) \sim r$, which will lead to a filament with constant mass per
unit length.  If we confine the starting sphere to a (narrow) range of
polar angles $\theta_1 \le \theta \le \theta_2$, the resulting surface 
density can be written in the form 
\be 
\sigma (\varpi) = {M_s \over 2 \pi \varpi^{3/2} \rot^{1/2} } 
\int_{x_2}^{x_1} {dx \over x^4} \bigl[ 1 - x^{-4} \bigr]^{-1/2} 
\, \ee
where $x_1 = (\sin\theta_1)^{-1/2}$ and $x_2 =
(\sin\theta_2)^{-1/2}$. Notice that we can confine the initial state
in azimuthal angle to a narrow range without affecting result (up to
the leading coefficient). In the limit that the range of polar angle
is also small, $\theta_2 = \theta_1 + \delta\theta$, the integral in
the above expression can be evaluated to obtain the result
\be 
\sigma (\varpi) = {M_s (\sin \theta)^{1/2} \delta\theta 
\over 4 \pi \varpi^{3/2} \rot^{1/2} } \, , 
\label{eq:sigfilone} 
\ee
where $\theta$ is the polar angle of the filament (so that $\theta_1
\to \theta$ and $\theta_2 \to \theta$; $\delta \theta \to 0$). Next we
want to write this expression for the surface density in terms of the
effective mass per unit length of the filament. The mass contained
within the boundary angles $\theta_1$ and $\theta_2$ is given by
\be 
M (r) = M_s (r / r_s) (\cos\theta_1 - \cos\theta_2) \, . 
\ee 
In the limit of interest, where $\theta_2 - \theta_1 = \delta \theta
\ll 1$, the effective mass per unit length $\mpl$ becomes 
\be 
\mpl = (M_s / r_s) \sin\theta (\delta \theta) \, . 
\ee 
The surface density can be written in the form 
\be 
\sigma (\varpi) = {\mpl r_s \over 
4 \pi \varpi^{3/2} (\rot \sin\theta)^{1/2} } \, . 
\label{eq:sigfiltwo} 
\ee

For comparison, we also consider a filament with mass per unit length
$\mpl$ with a polar angle $\theta$, where the mass in concentrated
along a single ray (centered on the origin).  Each point along the
filament, given by its radial coordinate $r$, has a particular angular
momentum given by $j = r^2 \Omega \sin\theta$.  As a result,
conservation of mass implies that each point along the filament
contributes to the surface density at a particular radius $\varpi$
within the disk, i.e.,
\be 
dm = 2 \pi \varpi \sigma d\varpi = \mpl dr \, . 
\label{eq:dfilament} 
\ee
Using the orbit solution from \S 2, we can relate the 
starting radial coordinate $r$ to the resulting disk radial 
coordinate $\varpi$, and find 
\be 
\varpi = \rot \sin\theta (r^2 / r_s^2) \, ,
\ee 
where $\rot = r_s \omega = r_s^2 \Omega / \sqrt{2\Psi_0}$. 
Using the above expression to evaluate $dr/d\varpi$, the 
surface density (see equation [\ref{eq:dfilament}]) can 
be written in the form 
\be 
\sigma (\varpi) = { \mpl r_s \over 4 \pi \varpi^{3/2} 
(\rot \sin\theta)^{1/2} } \, ,
\label{eq:sigfilray} 
\ee 
which is in agreement with the result of equation 
(\ref{eq:sigfiltwo}). 

As in previous collapse solutions, we can define a centrifugal radius
for filaments.  For the case considered here with constant mass per
unit length $\mpl$, this radius takes the form
\be 
R_C = (M_T / \mpl r_s)^2 \rot \sin\theta \, ,  
\ee 
where $M_T$ is the total mass that falls to the disk (from the
filament).  With this definition, the surface density becomes
\be
\sigma(\varpi) = {M_T \over 4 \pi 
\varpi^{3/2} R_C^{1/2} } \, \step (R_C - \varpi) \, . 
\label{eq:sigfilfinal} 
\ee 
The surface density thus takes a power-law form (here with index $q$ =
3/2) and has a sharp truncation radius at $R_C$. Notice that if we
integrate the surface density (eq. [\ref{eq:sigfilfinal}]) over the
disk, we obtain the expected disk mass $M_D$ = $M_T$ (mass
conservation is satisfied).

\section{MULTIPLE COLLAPSE EVENTS} 

One defining characteristic of the current picture of dark matter halo
formation is that mergers play an important role --- small halos come
together to forge larger bound structures. In most cases, some portion
of the gas in the merging units eventually falls into the center of
the composite structure, so that galactic disk formation must also
involve a series of ``accretion events''. The goal of this section is
to incorporate these multiple accretion episodes into our description
of gaseous disk formation (numerical simulations automatically include
this process -- see below).  Here we assume that the gas contributed
by each merging subhalo stacks up inside the gravitational potential
of the larger composite structure; after cooling, the gas falls inward
and thereby adds mass to the disk forming at the center.  The galactic
disk is thus assembled from inside-out, consistent with the standard
scenario (e.g., Murali et al. 2002, Maller et al. 2006).  Keep in mind
that the assimilation of a smaller halo does not provide a fully
formed disk, but rather provides a new source of gas for the composite
disk.  As a result, each accretion event produces a contribution to
the surface density of the composite disk. Further, this contribution
can be described as a disk surface density profile analogous to those
calculated above. To summarize, this section develops a description of
disk formation where a composite disk is assembled through the
addition of many separately accreted disk layers, where each layer is
provided by gas from a merging subhalo.\footnote{Although the net
effect of merging halos is to add disk surface density profiles
together, we stress that disks themselves are {\it not} merging.
Instead, the merging halos provide a new gas supply for continued
infall onto the disk. Notice also that we are making an assumption of
timing, i.e., that incoming halos have not yet formed their disks.}

Numerical simulations of hierarchical structure formation incorporate
the idea of multiple accretion events in a natural way. The addition
of mass into a developing halo is often described in terms of the halo
mass assembly history (MAH), which has been well studied (e.g.,
Avila-Reese et al. 1998; Firmani \& Avil-Reese 2000; Wechsler et al.
2002; van den Bosch 2002). This body of previous work indicates that
the MAH is generally not dominated by major mergers; instead the MAH
consists of a continuing series of smaller-scale events that can be
described as a smooth aggregation of mass (Murali et al. 2002; Maller
et al. 2006). In the (idealized) model developed herein, we consider
each of these small-scale accretion events to provide a supply of gas
that (after infall) results in a constituent surface density profile
that contributes to the overall composite disk profile. The smooth
nature of the MAH suggests that we must consider a large number of
accretion episodes (each with a relatively small baryonic mass). 

The distribution of angular momentum of the gas added during the
accretion events determines the distribution of centrifugal radii for
the constituent disks is thus an important quantity.  We expect the
newly added gas to have spin angular momentum inherited from the
rotation of the subhalo that supplied the gas. In addition, the
subhalo and its gas can have orbital angular momentum. Numerical
simulations indicate that galactic spin parameters $\lambda_j$ --- 
and hence the spin angular momentum of subhalos and their gas --- are 
drawn from a log-normal distribution (e.g., Bullock et al.  2001);
these numerical studies also suggest that the width of the
distribution $\sigwid \approx 0.5 - 0.6$.  Although the angular
momentum will include an additional orbital component, the latter
cannot dominate; otherwise, the total angular momentum of the disks
would be too large (see below). The distribution of orbital angular
momentum is not known, but it will tend toward a log-normal
distribution if a large number of physical variables play a role in
setting the orbits (due to the central limit theorem; Richtmyer 1978).
Another issue is that the angular momentum vectors (both spin and
orbital) of the constituents are not necessarily aligned. Since the
angular momenta add as vectors, this complication will act to make the
total (composite) angular momentum smaller (than the result of adding
scalars); further, this process adds another independent physical
variable which (through the central limit theorem) will help simplify
the composite distribution.  As a result, we expect the distribution
of total angular momentum to have nearly a log-normal form.

Before proceeding further, it is useful to summarize the ingredients
of the calculation. The validity of this approach relies on the
following assumptions: [1] The composite galactic disk forms as as
superposition of gaseous layers, where each layer can be considered as
a constituent disk with a power-law surface density distribution. We
first consider equal mass constituents and then generalize to varying
disk masses (which can be either correlated or uncorrelated with disk
radius).  [2] Each gaseous layer is assembled from the gas contributed
by a subhalo that merges with the growing galactic halo.  [3] During
the merging process, the gas of each ``accretion event'' retains its
component of angular momentum aligned with the overall sense of
rotation of the galaxy. [4] During the course of a particular
accretion event, infalling gas conserves its angular momentum and
produces a surface density profile in accordance with the solutions of
\S 2 -- 4.  The centrifugal radius of each disk layer is then
determined by the total (spin and orbital) angular momentum of its
progenitor subhalo.  [5] Over the course of multiple merger events,
the resulting distribution of (total) angular momentum is log-normal;
as a result, the distribution of centrifugal radii is also log-normal
with a well-defined peak value $R_0$ and width $\sigwid$.

This calculation begins with the simplest case of equal mass
constituent disks with the same general form, but with different
centrifugal radii (which correspond to to different angular momenta of
the pre-collapse gas).  The basic problem is thus to add together an
ensemble of contributing disk profiles with the form 
\be 
\sigma_j (\varpi) = {M_0 / N \over R_j^2} x^{-q} g(x) \, , 
\ee 
where $x = \varpi/R_j$ and $R_j$ is the centrifugal radius of the
$jth$ constituent disk (from the $jth$ accretion or merger event). 
The function $g(x)$ defines the shape of the edge, so that $g(1) \to
0$.  The analysis given above shows that typical edge functions are
$g(x) = \cos^{-1} x$ and $g(x) = (1-x^2)^{1/2}$.  The integer $N$
represents the number of events, and hence the total mass of the
composite disk is specified by $M_0$ times a dimensionless factor
given by the integral $I = 2 \pi \int_0^1 x^{1-q} g(x) dx$.

\begin{figure}
\figurenum{10}
{\centerline{\epsscale{0.90} \plotone{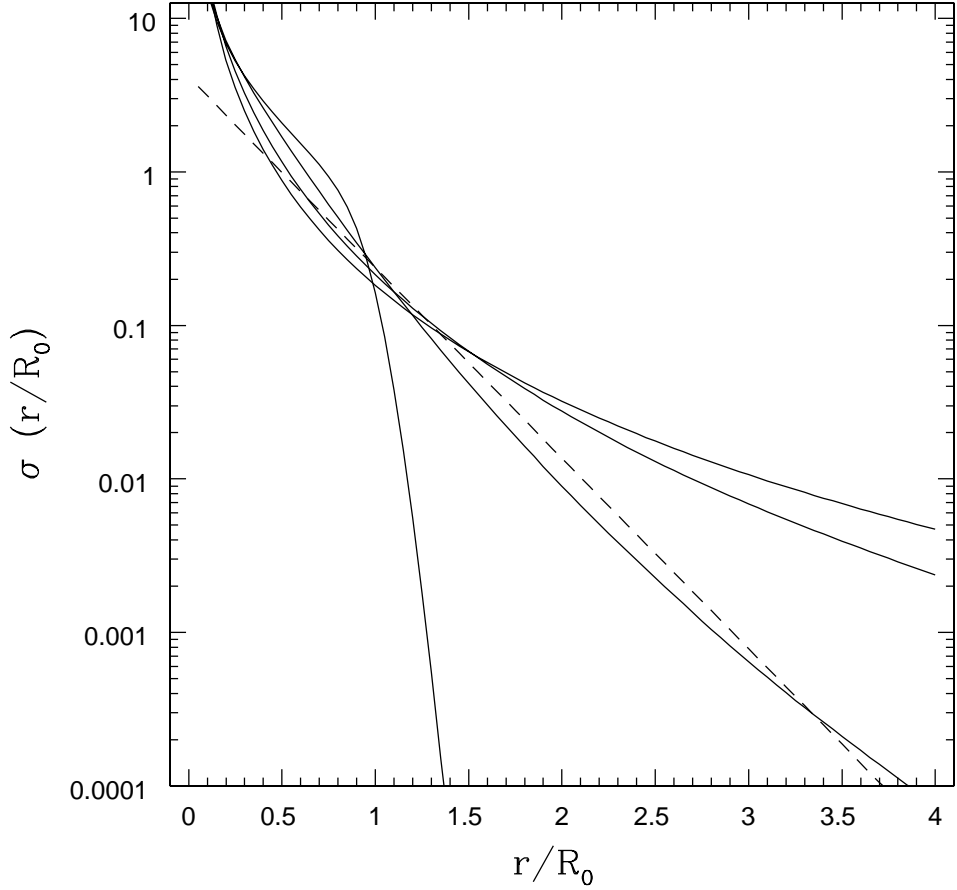} }}
\figcaption{The composite disk surface density profile resulting from
the addition of a large number of constituent disks. Each component
disk has the form $\sigma \sim \varpi^{-1} \cos^{-1}(\varpi/R_j)$,
i.e., a power-law with a well-defined outer edge at the centrifugal
radius $R_j$. The radii $R_j$ are drawn from a log-normal distribution
with varying dimensionless widths $\sigwid$ = 0.1 (lowest curve), 0.5,
1.0, and 1.5 (top curve). The radial scale is set by the peak of the
distribution at $R_0$. The dashed curve shows a reference exponential
profile, where the scale length $\rexp = 0.35 R_0$.  Notice that the 
intermediate value of $\sigwid$ = 0.5 comes close to producing an
exponential profile. Notice also that the scale on the vertical 
axis is arbitrary. }
\label{fig:comp} 
\end{figure}

In our formalism, the above considerations imply that the ensemble of
centrifugal radii are drawn from a log-normal distribution.  Let $R_0$
be the radius at the peak of the distribution and let $\sigwid$ denote
the width of the distribution.  The composite disk surface density
then takes the form
\be 
\sigma_T (\varpi) = {M_0 \over N} 
\sum_{j=0}^N {1 \over R_j^2} \Bigl( {R_j \over \varpi} 
\Bigr)^q g(x) \, . 
\label{eq:compsum} 
\ee
The sum can be evaluated by random sampling to determine the composite
surface density profile of the disk. One example is shown in Figure
\ref{fig:comp}, which uses the surface density distribution of
equation (\ref{eq:hqsigma}) to specify the constituent profiles. The
sum (\ref{eq:compsum}) is evaluated in the large $N$ limit. As 
expected, the composite surface density profile depends on the width
of the log-normal distribution. For small values of the distribution
width $\sigwid$, all of the components are nearly the same and the
composite profile looks much like that of the original constituents.
In the limit of large width $\sigwid$, the composite profile
approaches a power-law form ($\sigma \sim \varpi^{-2}$, as derived
below) over the range of radii of interest.  In the realm of
intermediate $\sigwid$, the composite surface density profile is close
to the exponential reference profile (shown by the dashed curve in
Fig. \ref{fig:comp}).  All of these results are robust (in that they
are largely independent of the assumed constituent profiles) and can
be understood analytically, as discussed below.

\begin{figure}
\figurenum{11}
{\centerline{\epsscale{0.90} \plotone{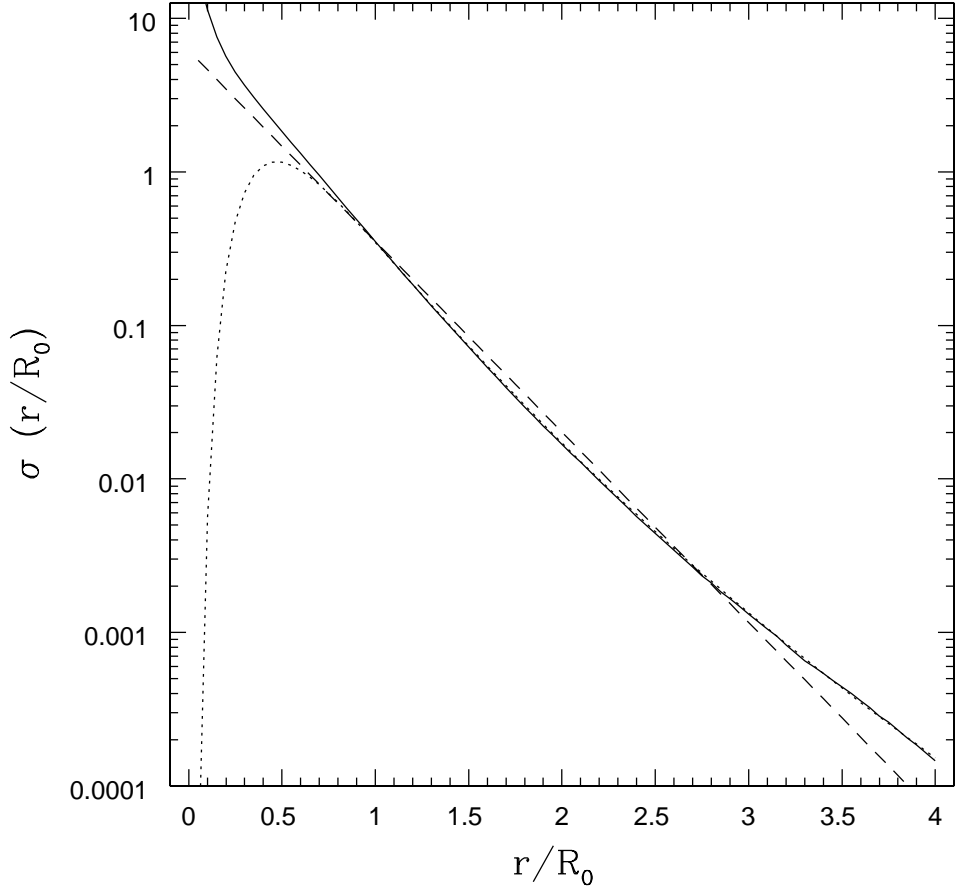} }}
\figcaption{The composite disk surface density distribution obtained
from random sampling compared to that obtained from the asymptotic
approximation derived in the text. The solid curve shows the composite
profile obtained by sampling $N$ = $10^5$ constituent surface density
profiles, with simple power-law form $\sigma \sim 1/ \varpi$,
truncated at centrifugal radii $R_j$, which are sampled from a
log-normal distribution of width $\sigwid$ = 0.5 (the radial scale is
set by the peak $R_0$ of the distribution of centrifugal radii). The
dotted curve shows the composite surface density obtained from the
approximation of equation (\ref{eq:approx}), which provides a good 
representation for $r \gta 0.6 R_0$. The dashed curve shows an 
exponential profile for comparison. The scale on the vertical 
axis is arbitrary. } 
\label{fig:comptest} 
\end{figure}

To study the problem of constructing composite surface density
profiles, we need to evaluate the sum in equation (\ref{eq:compsum}).
We consider the limiting case in which the disk edges are ``sharp'',
so that the edge function is a step function $g(x) = \step (1-x)$. For
a given realization of the disk parameters, we then order the disk
radii $R_j$, from largest to smallest, and define $F(\varpi)$ to be
the fraction of constituent disks with $\varpi < R_j$. The sum then
becomes
\be 
\sigma_T (\varpi) = {M_0 \over N} 
\sum_{j=0}^{F(\varpi)N} {1 \over R_j^{(2-q)} \varpi^q} \, . 
\label{eq:discretesum} 
\ee 
Since the $R_j$ are log-normally distributed, we can convert the sum
to an integral in the limit of large $N$, where a large number of
constituent disks are being assembled: 
\be
\sigma_T (\varpi) = M_0 \, \varpi^{-q} R_0^{-(2-q)} 
\int_{z_0(\varpi)}^\infty {dz \over \sqrt{2 \pi} \sigwid} 
{\rm e}^{-z^2/2 \sigwid^2} {\rm e}^{-(2-q)z} \, , 
\label{eq:original} 
\ee
where $z_0$ = $\ln (\varpi/R_0)$ and $z = \ln (R_j/R_0)$ is the
natural logarithm of the centrifugal radii being integrated over.
Recall that $R_0$ is a radial scale corresponding to the peak of the
log-normal distribution and $\sigwid$ is the distribution width.
Because the sum (eq. [\ref{eq:discretesum}]) only includes disks with
outer radii $R_j > \varpi$, the corresponding integral (eq.
[\ref{eq:original}]) has a limited range of integration from
$z_0(\varpi)$ out to $z \to \infty$. This integral can be written in
terms of error integrals, which have a number of known properties
(Abramowitz \& Stegun 1972; hereafter AS72). In particular, we find
\be
\sigma_T (\varpi) = M_0 \varpi^{-q} R_0^{-(2-q)} \, 
{\rm e}^{(2-q)^2 \sigwid^2/2} {1 \over \sqrt{\pi}} 
\int_{\xi_0}^\infty d\xi \, {\rm e}^{-\xi^2} \, , 
\label{eq:intstart} 
\ee
where $\xi_0 \equiv [z_0 + (2-q) \sigwid^2]/(\sqrt{2}\sigwid)$. 
To approximate the integral, we make use of the well-known inequality, 
\be
{1 \over \xi + \sqrt{\xi^2 + 2} } \le \, \, 
{\rm e}^{\xi^2} \int_\xi^\infty {\rm e}^{-t^2} dt \le 
{1 \over \xi + \sqrt{\xi^2 + 4/\pi} } \, , 
\label{eq:inequality} 
\ee
which is valid for $\xi \ge 0$ (AS72). This constraint allows us to write 
the composite surface density profile in the form 
\be 
\sigma_T (\varpi) \approx { M_0 \over \sqrt{\pi} \varpi^2} 
{ {\rm e}^{-z_0^2/2\sigwid^2} \over 
\xi_0 + \sqrt{\xi_0^2 + \const } } \, , 
\label{eq:approx} 
\ee
where both $z_0$ and $\xi_0$ are functions of the disk radial
coordinate $\varpi$.  The constant $\const$ must lie in the range
$4/\pi \le \const \le 2$, and can be chosen to minimize the error. If
we take $\const$ = 3/2, the approximation to the integral in equation
(\ref{eq:inequality}) is accurate to $\sim10\%$ for all allowed values
of the parameters, and is accurate to $\sim2\%$ over the range of
values needed to evaluate the surface density profile. Notice that
this form is only valid for $\xi_0 \ge 0$, which implies $\varpi \ge
R_0 \exp[-(2-q)\sigwid^2]$. For sufficiently large $\sigwid$, however, 
this form is valid over most of the radial range of interest. These
trends are illustrated in Figure \ref{fig:comptest}, which shows the
composite surface density calculated from random sampling compared to
that obtained using the approximation of equation (\ref{eq:approx}).

Now consider the composite distribution in the limiting cases. In the
limit where the width $\sigwid$ of the distribution is narrow, the
normal distribution (in the logarithm) becomes a Dirac delta function
(see eq. [\ref{eq:original}]) and we recover the expected result 
\be 
\sigma_T = {M_0 \over R_0^2} \Bigl( 
{R_0 \over \varpi} \Bigr)^q \, \step (1 - \varpi/R_0) \, . 
\label{eq:sigzero} 
\ee 
In the opposite limit of large width $\sigwid$, equation
(\ref{eq:approx}) is valid for essentially all radii as long as $q<2$,
and we find that the composite surface density takes a power-law form, i.e.,
\be 
\sigma_T \to {M_0 \over \sqrt{2 \pi} \sigwid (2-q)} \varpi^{-2} \, . 
\ee 
This resulting power-law form $\sigma_T \propto \varpi^{-2}$ is
independent of the index $q$ of the constituent disk profiles, as long
as $q < 2$. Notice that the results of \S 3 show that $q < 2$ for
essentially all cases of physical interest, as indicated by equation
(\ref{eq:qgeneral}).  For large values of the initial index $q > 2$,
all of the mass is concentrated toward the galactic center and the
composite surface density profile always takes the form $\sigma_T
\propto \varpi^{-q}$. This result follows from equation
(\ref{eq:intstart}).  As $\sigwid \to \infty$, $\xi_0$ becomes large
and negative, the integral in equation (\ref{eq:intstart}) approaches
a constant value ($\sqrt{\pi}$), and the composite surface density
takes the form $\sigma_T \sim \varpi^{-q}$.

In the limiting case of $\sigwid \to 0$, the disk edge becomes sharp,
which (in a sense) is much steeper than an exponential falloff (at
large $\varpi$). In the opposite limit of large $\sigwid$, the disk
has a power-law form $\sim \varpi^{-2}$, which falls off more slowly
than an exponential function. One might suspect that for some
intermediate value of $\sigwid$, the disk would obtain a roughly
exponential shape, at least for some range of radii. This behavior
does in fact occur, as shown in Figure \ref{fig:comp}, and can be 
understood in terms of the analytic approximation derived above. 
For radii sufficiently far from the origin so that equation
(\ref{eq:approx}) is valid, we can define an effective scale length
$\length$ for the surface density profile via
\be
\length^{-1} \equiv - {1\over \sigma_T} {d\sigma_T \over d\varpi} = 
{1 \over \varpi} \Biggl\{ 2 + {z_0 \over \sigwid^2} + 
\Bigl[ \bigl( z_0 + (2-q) \sigwid^2 \bigr)^2 + 2 \sigwid^2 
\const \Bigr]^{-1/2} \Biggr\} \, . 
\ee 
If this scale length $\length(\varpi)$ were exactly constant, when
considered as a function of $\varpi$, the surface density profile
would be exponential. In practice, this function is slowly varying.
Specifically, over the range of radii $1 \le \varpi/R_0 \le 5$ (which
corresponds to $0 \le z_0 \le 1.609$), the scale length $\length$
varies by 10--12\% for $q$ in the range 0.5 -- 1.5, where the width
$\sigwid$ is chosen to minimize the variation.  This optimization
requires values of $\sigwid$ = 0.3 -- 0.5, which are comparable to,
but somewhat smaller than, the widths of the distributions deduced
from numerical simulations (where $\sigwid \approx 0.5 - 0.6$; 
Bullock et al. 2001). 

The discussion thus far assumes that each constituent disk has a
constant mass (equal to $M_0/N$). However, the formalism is
sufficiently robust to consider the general case of varying disk
masses.  The composite profile will not change as long as the disk
masses are uncorrelated with the centrifugal radius $R_j$ of the
individual components.  Specifically, suppose that the mass of each
component disk has the form
\be
M_j = M_0 f_j (R_j/R_0)^\alpha \, , 
\ee
where $f_j$ is a random variable with unit mean. The masses are
uncorrelated when $\alpha$ = 0 and show a correlation with the
centrifugal radius for $\alpha \ne 0$. If we propagate this ansatz 
through the formalism given above, the composite disk profile takes 
the form 
\be 
\sigma_T (\varpi) = 
{ M_0 \over \sqrt{\pi} \varpi^{(2-\alpha)} R_0^\alpha } 
{ {\rm e}^{-z_0^2/2\sigwid^2} \over 
\xi_0 + \sqrt{\xi_0^2 + \const } } \, , 
\label{eq:genapprox} 
\ee
where the variable $\xi_0 \equiv [z_0 + (2-q-\alpha)
\sigwid^2]/(\sqrt{2}\sigwid)$.  Notice that the effect of the random
variable $f_j$ is completely washed out in the limit of a large number
of accretion events.  In the limit of a narrow distribution, we again
recover the expected result that the composite distribution is the
same as that of the components (eq. [\ref{eq:sigzero}]). In the limit
of a wide distribution, the composite surface density profile becomes
a power-law with index $2-\alpha$, independent of the index $q$ of the
component profiles, as long as $q < 2 - \alpha$. For ``large'' values
of $q > 2 -\alpha$, the composite surface density takes the form of
the original profiles so that $\sigma_T \sim \varpi^{-q}$.

\begin{figure}
\figurenum{12}
{\centerline{\epsscale{0.90} \plotone{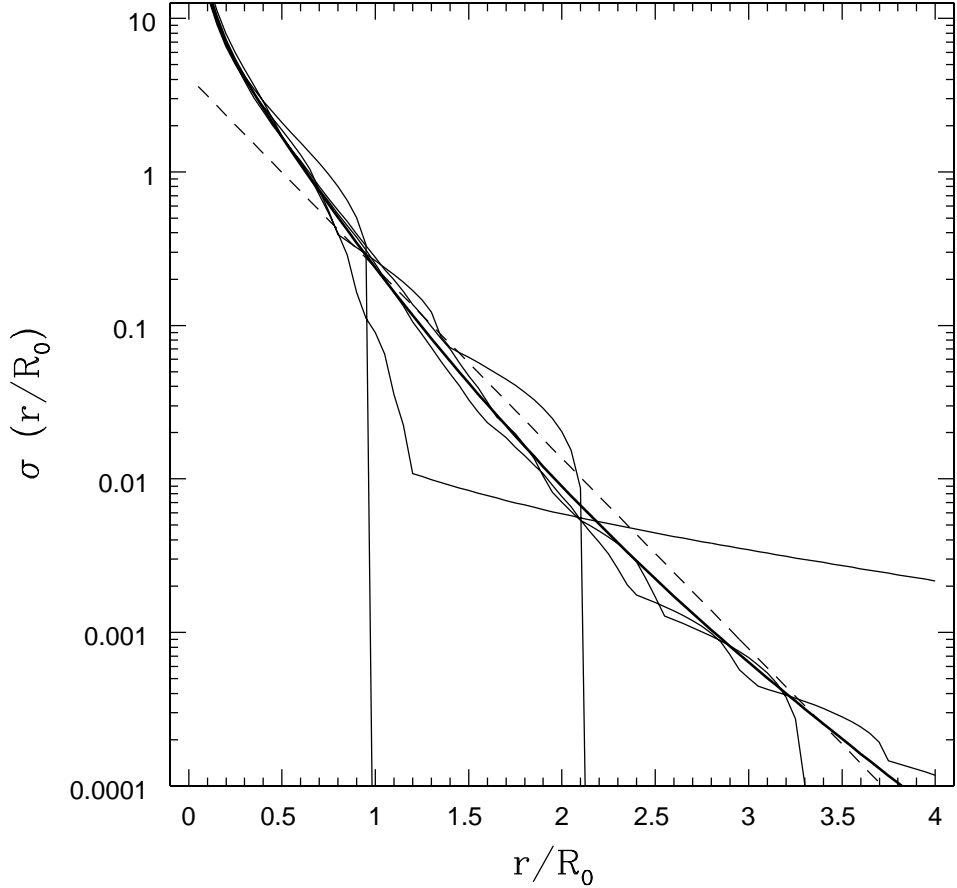} }}
\figcaption{The composite disk surface density profile for varying
number $N$ of constituent disks. Each component disk has the form
$\sigma \sim \varpi^{-1} \cos^{-1}(\varpi/R_j)$, where the radii $R_j$
are drawn from a log-normal distribution with dimensionless width
$\sigwid$ = 0.5 (and peak at $R_0$).  The heavy solid curve shows the
resulting profile in the limit of large $N$; the lighter solid curves
show the results for $N$ = 1, 4, 16, 64, and 256, where the curves
approach that of the large $N$ limit with increasing $N$. Notice that 
for small $N$, the profile will vary from realization to realization 
due to incomplete sampling of the distribution of disk radii. The 
dashed curve shows an exponential surface density profile as a 
reference point. The scale on the vertical axis is arbitrary. } 
\label{fig:compvsn} 
\end{figure}

The results presented thus far have been calculated in the limit of
large $N$, so that the surface density of the composite system
approaches a particular smooth form. An interesting question is to ask
how large $N$ must be in order for the problem to be safely in the
large $N$ limit.  One way to answer this question is to calculate the
RMS deviation of the difference between the composite surface density
obtained with a ``small'' value of $N$ and that of the large $N \to
\infty$ limit.  Notice, however, that for small $N$, each realization
of the random values (the $R_j$) will lead to a somewhat different
profile and hence a different measure of the RMS error. Roughly
speaking, for $N$ = 10, we get deviations of about 50 percent, for $N$
= 100 the deviations are about 10 percent, and for $N$ = 1000 the
deviations are only about 2 percent. These deviations are computed
over the range $0 \le \varpi \le 2 R_0$; they are greater at larger
radii.  Another way to address this issue is to plot the resulting
surface density profiles for varying numbers $N$ of the constituent
disks. The result is shown in Figure \ref{fig:compvsn} for
logarithmically spaced $N$ = 1, 4, 16, 64, and 256.  As the number of
constituent disks $N$ increases, the profiles approach that of the
asymptotic (large $N$) limit shown by the heavy solid curve in Figure
\ref{fig:compvsn}.  Keep in mind that the profiles will vary greatly
from realization to realization for small values of $N$; in other
words, the small-$N$ profiles shown in the figure will look different
every time they are computed (with independent sampling).
Nonetheless, for $N \gta 16$, the composite surface density has
roughly the same form as that of the asymptotic limit out to $r
\approx 2 R_0$. Similarly, for $N \gta 64$, the surface density
profile will be similar to the asymptotic limit out to $r \sim 3 R_0$.
This result is important, since galaxy sized halos are thought to be
assembled from moderate numbers of component halos (Wechsler et
al. 2002).

This formalism can be applied to any scenario in which the net effect
is to add surface density profiles together. As one example, it can
account for cosmological evolution of the background potential. In the
approximation used here, the surface density profiles of the
constituent disks add together so that the final state does not depend
on the order in which they are assembled.  In a cosmological setting,
however, dark matter halos become steadily larger, both from continued
infall from large distances and through the mergers of smaller
halos. As a result, the depth of the gravitational potential well
($\Psi_0$) increases with time. If the constituent disks result from
the collapse of gas that attains hydrostatic equilibrium (\S 3.5),
then the pre-collapse extent of the gas will be smaller at later times
(for a fixed mass), and the corresponding centrifugal radius $R_C$
will be a decreasing function of time if angular momentum is constant.
On the other hand, some observations (Elmegreen et al. 2005) and
theoretical work (Firmani \& Avila-Reese 2000) indicate that disk
radii should increase with time. This result can be made consistent
with our multiple accretion scenario if the specific angular momentum
increases for subsequent accretion events. In any case, the time
dependence of the centrifugal radii can be taken into account in the
distribution of $R_j$. If the distribution is log-normal, as assumed
here, then the final result depends only on the peak value $R_0$ and
width $\sigwid$ of the distribution. However, this procedure can be
generalized to other distributions. As another example, some numerical
simulations suggest that baryons can transfer angular momentum to the
dark matter so that gas falling in later would have a systematic shift
in its angular momentum distribution (private communication, anonymous
referee); the effect of this process on the resulting surface density
profile of the disk could be studied with the formulation developed
here.

This analysis shows that a composite disk can display a nearly
exponential surface density profile, provided that the distribution of
centrifugal radii has an intermediate width ($\sigwid \approx 0.5$)
and the constituent disks have power-law indices $q < 2$. Both of
these conditions are expected to be met, as outlined above. Notice
that when an exponential surface density profile is realized, the
scale length is roughly $\rexp \approx 0.35 R_0$ (Figs. \ref{fig:comp}, 
\ref{fig:comptest}, \ref{fig:compvsn}), where $R_0$ is the centrifugal
radius at the peak of the distribution. To produce scales $\rexp
\approx 4 - 5$ kpc, as indicated by observations (Binney \& Merrifield
1998), the radius $R_0$ must lie in the range $R_0$ = 11 -- 14 kpc.
The centrifugal radius of an individual constituent disk depends on
the spin parameter $\lambda$ through a relation of the form $R_C =
\alpha \lambda r_s$, where $\alpha$ is a dimensionless parameter of
order unity. For example, if the halo has a Hernquist profile, the gas
density follows that of the dark matter, and all of the gas out to $r
= r_s$ falls in, then $\alpha \approx 4$.  With typical halo scale
radii $r_s$ = 50 -- 70 kpc and spin parameters $\lambda$ = 0.035 --
0.04 (at the peak of the distribution -- see Bullock et al.  2001 and
Wechsler et al. 2002), the expected peak value of the centrifugal
radius would lie in the range $R_0$ = 7 -- 11 kpc. These values of
$\lambda$ are thus comparable to, but somewhat lower than, those
required to produce the correct disk scale length.  However, the
values quoted above only include the spin angular momentum of the
constituent gas modules. As these subunits fall into the larger halo
and merge, and thereby add their gas to the growing structure, they
contain orbital angular momentum in addition to that described by the
spin parameter $\lambda$.  If the orbital angular momentum is roughly
comparable to (but smaller than) the spin angular momentum, the
constituent disks will have the proper total angular momentum for this
multiple infall process to produce exponential composite disks with
the observed scale length.

\section{CONCLUSION} 

\subsection{Summary of results} 

This paper has studied the assembly of galactic gaseous disks by
following orbit solutions for baryonic gas parcels as they fall
through the potential wells of dark matter halos. Our general analytic
formulation of the problem (as presented in \S 2) is robust and can be
used in variety of contexts. In this paper, we calculate the velocity
fields of the incoming gas, the density distribution of the gaseous
material as it falls inward, and the surface density profiles of the
resulting gaseous disks.  Our specific results are summarized below:

[1] The generic outcome for a gaseous disk produced by a single
spherical collapse is a power-law surface density that is truncated at
a centrifugal radius $R_C$ (see Figs. \ref{fig:profiles},
\ref{fig:profiles2}, and \ref{fig:isodisk}). Here, the centrifugal
radius $R_C$ is the radius to where the material with the highest
(initial) specific angular momentum falls during the collapse. This
result suggests that the ``fundamental units'' of galactic disk
assembly are not exponential disks, but instead are truncated
power-laws.  For a constant rotation rate of the initial gaseous
sphere, the power-law index $q$ of the surface density distribution
($\sigma \propto \varpi^{-q}$) depends on the initial mass
distribution such that $q = 2 - p$, where $p$ is the power-law index
of the starting mass profile ($M \propto r^p$). For a rotation rate
that varies with radius according to $\Omega \sim r^{-s}$, the
power-law index of the resulting disk surface density is given by $q =
2 - p/(2-s)$. We also determined the density profiles appropriate for
hydrostatic equilibrium of both isothermal and polytropic gas in the
background potential of a dark matter halo (\S 3.5) and followed the
subsequent collapse (Figs. \ref{fig:isodisk} and \ref{fig:polytrope}).
For polytropic equations of state with relatively large indices $n$,
the resulting surface density profiles are close to exponential
(except near the centrifugal barrier). 

[2] In addition to determining the disk surface density profiles, this
calculation provides an analytic description of the dynamics that
builds up galactic disks. This description includes the velocity
fields of the gas during the collapse (eq. [\ref{eq:vfield}]), the
density distribution of the infalling baryonic material
(eqs. [\ref{eq:density0}] and [\ref{eq:density1}]), and the mapping
between the starting locations in the pre-collapse cloud and the
``final'' radial positions within the disk (e.g., Figs.
\ref{fig:diagram} and \ref{fig:conediagram}). We also obtain analytic
forms for the hydrostatic equilibrium profiles of polytropic gaseous
spheres in the gravitational potential of a dark matter halo
(eqs. [\ref{eq:isodense}] and [\ref{eq:polydense}]). 

[3] This collapse picture has been generalized to include less
idealized starting configurations (\S 4). The first generalization
considers the starting gaseous spheres to have evacuated regions in
their centers. The signature of these central holes in the starting
states is to produce impoverished central regions in the corresponding
gaseous disks (see Fig. \ref{fig:radholes}). The starting states can
also be considered to have limited angular extent. The removal of the
polar caps from the initial states also leads to a deficit of material
in the disk centers (relative to the spherical initial state) as shown
in Figure \ref{fig:polarholes}.

[4] This formulation of the collapse problem can be further
generalized to include multiple accretion events, which are expected
to occur due to merging of individual halos to form larger, composite
structures (\S 5). Each incoming halo provides a new supply of gas,
which subsequently cools and falls inward to produce another component
to the disk.  Here we consider the addition of a large number of such 
constituent disks, where the centrifugal radii are drawn from a
log-normal distribution (consistent with the numerical finding that
halo spin parameters display a log-normal distribution) and find an
interesting and somewhat unexpected result: The addition of a large
number of power-law disk profiles with index $q < 2$ results in a
composite surface density profile of the form $\sigma \sim
\varpi^{-2}$ in the limit of a wide distribution for $R_j$, where this
result is independent of the index $q$ (provided that $q < 2$). In the
limit of a narrow distribution of $R_j$, the composite surface density
is the same as the constituent surface density profiles. In the
intermediate regime, however, the composite surface density can
display a nearly exponential form over a range of radii (Figs.
\ref{fig:comp} and \ref{fig:comptest}). The asymptotic forms are
obtained in the limit of large numbers $N$ of constituent disks, and
the approach to large $N$ is shown in Figure \ref{fig:compvsn}.
Although the large $N$ limit is approached rather slowly, this
sequence shows that tens of subunits (constituent disks) are necessary
to produce a smooth disk with a nearly exponential surface density
profile.

\subsection{The quest for exponential disks} 

One of the defining features of observed disk galaxies is that their
surface brightness distributions display a nearly exponential form.
One might hope that numerical simulations of galaxy formation would
naturally produce gaseous disks with exponential surface density
profiles, so that star formation could take place {\it in situ} and
thereby account for the observations. Unfortunately, the production of
exponential surface density profiles is not automatic. Building upon
earlier work (Dalcanton et al. 1997, van den Bosch 2001, Kaufmann et
al. 2006, and many others), this paper shows that the fundamental
disks produced by a single collapse event have the form of a truncated
power-law rather than an exponential profile (Figs. \ref{fig:profiles}
and \ref{fig:profiles2}). The transformation between these truncated
power-law forms and exponential profiles most likely involves several
processes, as outlined below:

[A] The observed disk brightness profiles are close to exponential
only over a limited range of radii (roughly 2 -- 20 kpc) and a
correspondingly small range of scale lengths (roughly $r/\rexp \approx
0.5 - 5$). The inner regions depart from an exponential form due to
contamination from bulge components and the outer regions are too
faint to be observed. Further, as shown by Figure \ref{fig:profiles},
the difference between a truncated power-law and an exponential is not
too large over a limited range of radii. One of the largest
differences between the power-law profiles and an exponential profile
occurs at small radii, where the power-law profiles have cusps; this
discrepancy is largely alleviated, or at least masked, by the presence
of the bulge component (for $r < 2$ kpc). In addition, if the initial
(pre-collapse) gas departs from a complete spherical form, either
through inner evacuated regions or the removal of polar caps, then the
resulting surface density profiles will not have inner cusps (Figs.
\ref{fig:radholes} and \ref{fig:polarholes}), or they will be less
pronounced.

[B] The calculation presented in this paper accounts for only the
formation of the disk, not its subsequent evolution. If galactic disks
are produced with sharp edges (as indicated here for the case of a
single collapse event), then gravitational instabilities will grow
robustly and cause the disk to spread.  Although previous authors
(Dalcanton et al. 1997) have argued that the time scale for such
instability is too long, the quoted (long) time scales are appropriate
for disks that already have soft (e.g., exponential) edges. If disks
have sufficiently hard edges, as produced here, then gravitational
instabilities can grow on the dynamical time scale of the outer disk
edge (Adams et al. 1989; Ostriker et al. 1992). This time scale is
given by the inverse of the orbital frequency at the centrifugal
radius, roughly $\tau = R_C / v_{rot} \approx$ 100 Myr ($R_C$/20
kpc). This time scale is still somewhat long to account for all of the
necessary angular momentum redistribution, but gravitational
instabilities will have time to grow and act to spread out the
disk. Notice that a truncated power-law profile that spreads out will
become more like an exponential profile.
 
[C] Exponential profiles are observed for the surface brightness
distributions of galactic disks, not the surface density of the gas,
so it is possible for the star formation process to help produce the
observed exponential forms. A good demonstration of this effect is
found in Lin \& Pringle (1987), where initial surface density profiles
roughly like those of Figure \ref{fig:profiles} spread out under the
action of disk viscosity and convert some fraction of the mass into
stars.  The resulting surface density of the stars is much more like
an exponential form than the starting gas profile.

[D] Finally, as shown in \S 5, the addition of multiple disk surface
density profiles can produce a composite profile that displays a
nearly exponential form --- much more like an exponential than the
initial (constituent) starting profiles. The ingredients necessary to
produce a nearly exponential profile are a large number of constituent
disks and a log-normal distribution for their centrifugal radii. This
result is independent of the shape of the constituent disk profiles as
long as they have effective power-law indices $q < 2$.  Since the
current paradigm of structure formation indicates that galaxies are
assembled through a process of hierarchical merging, multiple
accretion events are highly likely. Further, numerical simulations
suggest that the distribution of halo angular momentum is log-normal
(Bullock et al. 2001) so that the distribution of centrifugal radii is
also expected to be log-normal.

In practice, all four of these considerations (and perhaps others)
will conspire to produce the observed exponential disk profiles.
Individual infall events will each produce disk surface density
profiles that are truncated power-laws, but multiple infall events
will jointly build up a nearly exponential disk (as shown in Fig.
\ref{fig:comp}). The effects of star formation, viscous spreading, and
disk evolution via gravitational torques will act to smooth out the
disk and drive the surface density towards a more exponential form.
Given the presence of the bulge contribution at small radii and the
limited dynamic range of the observed profiles, the exponential disks
in observed galaxies can be understood through the joint action of
these processes.


\newpage 

\centerline{\bf Acknowledgment} 

We would like to thank Michael Busha, Gus Evrard, and Lars Hernquist
for many beneficial discussions.  This work was supported at the
University of Michigan by the Michigan Center for Theoretical Physics;
by NASA through the Terrestrial Planet Finder Mission (NNG04G190G) and
the Astrophysics Theory Program (NNG04GK56G0); and by NSF through
grants DMS01038545 and DMS0305837. Finally, we thank the referee for a
comprehensive and detailed set of comments that improved the manuscript.

\end{document}